\documentstyle[preprint,tighten,aps,12pt]{revtex}
\begin{document}
\bibliographystyle{unsrt}
\preprint{UBCTP-93-20}
\title{
Tunneling in Quantum Wires: a Boundary Conformal Field Theory Approach
}
\author{Eugene Wong$^b$ and Ian Affleck$^{a,b}$}
\address{
$^a$Canadian Institute for Advanced Research and $^b$Department of
Physics,\\ The University of British Columbia, \\ 6224 Agricultural
Road, Vancouver, B.C. Canada V6T 1Z1 }
\maketitle
\def\tq{{\tilde q}}
\def\tn{{\tilde n}}
\def\bcp{boundary critical phenomenon}
\def\tphi{{\tilde \phi}}
\def\mod{{\rm mod}}
\def\Qh{{\hat Q}}
\def\Ph{{\hat \Pi}}
\def\tpL{{t+x \over L}}
\def\tmL{{t-x \over L}}
\def\pa{\partial}
\def\>>{\rangle}
\def\<<{\langle}
\begin{abstract}
Tunneling through a localized barrier in a one-dimensional interacting
electron gas has been studied recently using Luttinger liquid
techniques.  Stable phases with zero or unit transmission occur, as
well as critical points with universal fractional transmission whose
properties have only been calculated approximately, using a type of
``$\epsilon$-expansion''.  It may be possible to calculate the
universal properties of these critical points exactly using the recent
boundary conformal field theory technique, although difficulties
arise  from the $\infty$ number of conformal towers in this $c=4$
theory and the absence of any apparent ``fusion'' principle.  Here, we
formulate the problem efficiently in this new language, and recover the
critical properties of the stable phases.
 \end{abstract}
\pacs{  }
\section{Introduction}
The low energy transmission of a 1-d interacting electrons through
barriers, including
resonant tunneling, was analyzed by mapping the problem to a
Luttinger liquid where the gapless spin and charge degrees of freedom
decouple, interacting with a barrier or a quantum impurity\cite{Kane}
\cite{Furusaki}.  A simple lattice version of the underlying microscopic
Hamiltonian is given by the Hubbard model with a spin-spin interaction
which breaks the $SU(2)$ spin symmetry: \begin{eqnarray}
H &=& \sum_{i}\biggl[ \left(t_i \psi^{\alpha \dagger}_i\psi_{\alpha ,i+1}+
h.c.\right)+U\left(\psi^{\alpha \dagger}_i\psi_{\alpha ,i}\right)^2+
JS_i^zS_{i+1}^z \label{micHam}
+\mu_i \psi^{\alpha \dagger}_i\psi_{\alpha ,i} + h_iS_i^z \biggr] \\
&{\rm where}&~~ S_i^z=
\psi^{\alpha \dagger}_i\left( \sigma^3\right)_\alpha^\beta \psi_{\beta ,i}~.
\nonumber \end{eqnarray}
Here
the hopping amplitude $t_i$, chemical potential $\mu_i$ and magnetic
field $h_i$ are constant,  except in the vicinity of the
origin. The Fermi energy is arbitrary except that we stay away from
half-filling, so that both charge and spin excitations are gapless.
We
will often be interested in cases where there is parity symmetry (which
may be reflection about a site $\psi_i \rightarrow \psi_{-i}$ or about
a link $\psi_i \rightarrow \psi_{-i+1}$) that requires $t_i$ to be real
or where there is
symmetry of spin rotation about the $x$-axis by $\pi$: $S_i^z\to
-S_i^z$ which requires $h_i=0$. In the
simplest case of a local barrier, only $t_0$ differs from
the other $t$'s, and we set
$\mu_i$ constant and $h_i=0$.  In the
resonant tunneling case, we take $t_{-1}=t_0$ different from the other
$t$'s.  Now the site $0$ is distinguished, so we choose $\mu_0$
different than $\mu_i=\mu$ on all other sites.  By fine-tuning these two
parameters describing the double barrier, we can achieve resonances.
We emphasize
that we are concerned  with universal, low energy properties so that the
detailed form of the microscopic Hamiltonian is unimportant.

In  the
case of spinless
fermions scattering off a potential barrier, Kane and Fisher \cite{Kane}
showed that at zero temperature, the charge conductance is zero
if the {\it bulk} interactions are repulsive and perfect if
attractive.  More generally, for fermions with
spin, the charge and spin conductances depend on two parameters which
are related to the bulk interactions $U$ and $J$ of (\ref{micHam}) in
the charge and spin sectors.  It was  found
that there are four possible stable phases whose stability depend on the
strength of the bulk interactions: charge and spin with zero or perfect
transmission.    In addition, there exist unstable phases which
have partial conductances separating pairs of the above phases in the region
of overlap of the domains where the two phases are stable. These unstable
phases were probed perturbatively, using a type of
``$\epsilon$-expansion'' based on the observation that when the bulk
interaction constants approach certain values these fixed points become
trivial.  Our hope is that by using the recently developed boundary
conformal field theory
technique\cite{Afflecko,Affleckgsd,Affleckti,Affleckh,Ludwig,AffleckG,Eggert}
we can determine their properties non-perturbatively.   But before diving into
the nontrivial unstable boundary fixed points, we have to verify that this
method is applicable and that in the new formalism, it does reproduce all the
basic features of the problem.  This is the purpose of this paper.

The present problem fits into the more general setting where we have
one-dimensional gapless degrees of freedom in the bulk coupled to a
local potential or impurity degree of freedom.  Such systems have been
tackled by the boundary critical phenomenon approach in the Kondo problem
and in the isotropic spin-$1 \over 2$ antiferromagnetic
Heisenberg chain with an
impurity
\cite{Afflecko,Affleckgsd,Affleckti,Ludwig,AffleckG,Eggert}.
However,  in the Heisenberg chain after a Jordan Wigner
transformation, the bulk is composed of interacting
{\it spinless} fermions
and in the Kondo case, {\it free} spinful fermions.
Here we are interested in effects of local
interactions in an interacting spin-$1 \over 2$ gas of
fermions.

In general, the system of gapless degrees of freedom coupled to a
local degree of freedom is a difficult problem to solve exactly even
in 1-d.  There exists exact solution from the Bethe Ansatz but the
Hamiltonian must be fine tuned to become integrable \cite{Sorensen}.
For a generic situation, we simplify the problem by asking what the
low energy behavior of the system is.
At long wavelengths and low energies, we can describe the bulk by a
relativistic $(1+1)$-dimensional field theory with conformal invariance.
In the boundary critical phenomenon, we do not integrate out the bulk
degree of freedom as in \cite{Kane,Furusaki}, but propose that at low
energies, the effects
of local interactions with the barrier or the impurity can be
summarized by an effective boundary condition
on the bulk.   The boundary condition must renormalize to a fixed
point, so that it will be compatible with the bulk conformal symmetry.
At such a boundary fixed point, conformal symmetry in $(1+1)$-dimension
is powerful enough to give, for example, the finite size spectrum.
By turning the present problem into a boundary critical phenomenon, the
four stable phases and the unstable ones mentioned above will
correspond to the various conformally invariant
boundary conditions on the bulk.
As in the Kondo and Heisenberg chain, we will follow
Cardy's approach to boundary critical
phenomenon to treat the present problem \cite{Cardy89,Cardy86}.

There are however two aspects that are new to this problem that were
not present in Cardy's treatment nor in the Kondo or Heisenberg problem.
The first concerns the symmetry of the problem.
This eventually leads to irrational conformal field theories rather
than rational ones like the other cases.
Cardy concentrated only on
the $c<1$ conformal field theories.  Viewing the problems as
$(1+1)$-dimensional field theories, the states in the Hilbert space can be
classified into a finite number of primary states and an infinite number
of descendents.  For instance, in the $c={1 \over 2}$ Ising case, we have
three primary  states.  For the Kondo and Heisenberg problems, the bulk and
the spin-spin interactions with the impurities are both spin $SU(2)$
invariant. Although we have $c \ge 1$
conformal field theories, the extra $SU(2)$ symmetry enlarge the
symmetry group to $SU(2)$ Kac-Moody symmetry.  Once again, the
states in the Hilbert space can be classified into finite number of
primaries and the rest descendents.  A finite number of primary
states has the nice feature that the modular transformation
\cite{Ginsparg} needed in the
boundary critical phenomenon approach is linear and is given by a
finite-dimensional matrix, known as the modular $S$-matrix.  We
will make this point clear in the next section.
For the case at hand, we have a $U_C(1)\times U_S(1)$ symmetry for the
conservation of the fermion's charge and the $z$-component of the
spin. The interactions with
the local potential or the impurity are through both spin and
charge.  Any interactions must preserve this
$U_C(1)\times U_S(1)$ symmetry.
(In the special case when the bulk and boundary spin interaction
preserve rotational symmetry, then we recover $U_C(1)\times SU_S(2)$ symmetry.
For the spinless fermions, we only have the $U_C(1)$ symmetry.)
However, the $U(1)$ Kac-Moody symmetries are not restrictive
enough to group the spectrum of this $c \ge 1$ conformal field theory
into a finite number of conformal towers.  With an infinite number of
primaries, the modular transformation is given by an integral equation and
not a finite-dimensional $S$-matrix.   However, we are able to generalize
Cardy's approach since it is the partition  function that is important, not
the fact that the number of primaries is finite. A similar problem was
dealt with
recently in the boundary conformal field theory of monopole-catalyzed
baryon decay.\cite{Sagi}

The other somewhat new aspect has to do with the fact that
 Cardy's formalism assumes that the boundary conditions do not allow
momentum to flow across the boundary.  For instance, in Cardy's treatment of
the Ising model, the three conformally invariant boundary conditions are
spin up, down and free.  Here, in the extreme case when
the electrons are perfectly transmitting across
the barrier, we anticipate a fixed point at which all the
charge coming in is being transmitted across the boundary to
the other side.  To transform such a fixed point into a boundary fixed
point of Cardy's type, we fold the system at the boundary so that we
essentially turn the system into one defined on the half-line
by doubling the number of bulk degrees of freedom.  The same trick was used
in the boundary conformal field theory treatment of the two-impurity Kondo
effect \cite{Affleckti}.

The paper is organized as follows.
In section II, we generalize Cardy's \bcp\ approach to include the
present problem.
We first demonstrate how this is applied to the simpler case of
spinless fermions in section III.  In particular, we will give the
finite size spectra corresponding to the various conformally
invariant boundary conditions.  We will then give the conductance
formula and discuss the stability of these boundary fixed points by
examining the operator contents.
We will see that our results agree with \cite{Kane}.  We will also
calculate the ground state degeneracy, $g$ for periodic and open boundary
conditions, showing that the `$g$-theorem' \cite{Affleckgsd} is
obeyed. In section IV, we
extend our analysis for the spin-${1\over 2}$ fermions.  All our
results are in full agreement with Ref. \cite{Kane}, after
correcting a minor error in that work.  We conclude in section IV.

\section{Generalizations of  Cardy's approach}
We will recall briefly and generalize the ingredients of Cardy's
approach to \bcp\ \cite{Cardy89}.  Consider some conformally
invariant boundary
conditions imposed along the real axis of the complex plane $z=\tau +
ix$, giving the geometry of the upper-half plane.  The conformal field
theory is invariant under infinitesimal coordinate transformation $z
\rightarrow  \sum_n a_n z^{n+1}$.  In order to preserve the boundary
$x=0$, $a_n$ must be real.  Truncating half of the infinitely many
symmetry transformation leads to half as many conserved charges,
\begin{equation}
L_n={1 \over 2\pi i}\int_{C_+}z^{n+1}T(z)dz
-{1 \over 2\pi i}\int_{C_+}{\bar z}^{n+1}{\bar T}({\bar z})d{\bar z}
\label{evirasoroa}\end{equation}
where $C_+$ is a semicircle contour in the upper-half plane
and there are no ${\bar L}_n$'s.
Here $T$ and $\bar T$ are the left and right-moving components of the
Hamiltonian density.
For the Ward identity to continue to be valid, we impose
$T-{\bar T}=0$
along the real axis so that there is no contribution from the integral
along the real axis part of the contour $C_+$.  In other words, $T_{\tau
x}=0$ is imposed at the boundary, meaning no momentum flux across it.  This
condition allows us to think of ${\bar T}({\bar z})$ in the
upper-half plane as $T(z)$ in the lower-half plane, yielding
\begin{equation}
L_n={1 \over 2\pi i}\oint_C z^{n+1}T(z)dz \label{evirasorob}
\end{equation}
where $C$ is the circle at infinity.
We therefore have a purely holomorphic (left moving) system when we
extend from the
half-plane to the entire complex plane.
In addition to the conformal symmetry, we also have  $U(1)$
symmetries, corresponding to the charge and the $z$-component of spin.  For
a $U(1)$ current, $J(z)$, the analogous equations hold: \begin{eqnarray}
J_n&=&{1 \over 2\pi i}\int_{C_+}z^{n}J(z)dz
-{1 \over 2\pi i}\int_{C_+}{\bar z}^{n}{\bar J}({\bar z})d{\bar z} \\
&=&{1 \over 2\pi i}\int_Cz^{n}J(z)dz  \label{ekacm}
\end{eqnarray}
where $J(z)-{\bar J}({\bar z})=0$ along the real axis.
By mapping the upper-half plane to an infinite strip of
width $l$ by $w={l \over \pi}ln z =u+iv$ where $u$ and $v$ are the
coordinates for time and space respectively, one can show that
\cite{Cardy86}
\begin{equation}
H={1\over 2\pi} \int_o^l (T(w)+{\bar T}({\bar w}))dv = {\pi\over l}(L_0
-{c\over 24}) \label{ehamil}
\end{equation}
is the Hamiltonian on the strip.

The idea of Cardy for  rational conformal field theory can be
generalized to the present case since it is the partition function
that Cardy worked with.
Recall that imposing modular invariance ($S$ and $T$) on a conformal
field theory defined on the torus (that is periodic boundary
conditions imposed on the fields along both nontrivial cycles) determines
the operator content of
the bulk conformal field theory \cite{Cardyoc}.  Here, for the finite
size system
with boundaries, instead of a torus, we have the geometry of a
cylinder by making the strip periodic in time.  One can still
impose the $S$ modular invariance to
determine the operator content of the conformal field theory with
boundaries.  We will now recall how this is done and generalize this
process.

Let $H_{ab}$ be the Hamiltonian with boundary conditions $a$ and $b$
at the two ends of the finite strip on the complex plane
$w=u+iv$.  For a finite temperature field
theory, the Euclidean time $u$ is defined modulo
$\beta$, the inverse temperature. The partition function then is
given by
\begin{eqnarray}
Z_{ab}&=&Tr{\rm exp}(-\beta H_{ab})=\sum_i\chi_i(q)n^i_{ab} \label{epartab}\\
{\rm where} ~~
q&=&{\rm exp}(-{\pi\beta \over l})~~ {\rm and}~~
\chi_i(q)=q^{-{c \over 24}} Tr_i q^{L_0}. \label{echardef}
\end{eqnarray}
(\ref{ehamil}) has been used to derive the above equation and we are going to
determine $n^i_{ab}$, the
number of times that the $i^{th}$ conformal tower appears in the system
spectrum with boundary conditions $a$ and $b$ imposed at the two boundaries.
The trace in $\chi_i$ is over the descendent states of the $i^{th}$
primary.
In the $c=1$ $U(1)$ Kac-Moody theory, we will see that $\chi_{i=Q}$
are of the form
\begin{equation}
\chi_Q(a,\delta;q)
={1\over \eta (q)}
q^{{a\over 2}(Q-{\delta\over 2\pi})^2}
\label{echaracter}
\end{equation}
where $\eta(q)=q^{1\over 24}\prod_{k=1}^\infty(1-q^k)$ is the Dedekind
function, $Q$ is an integer, $a$ and $\delta$ are real parameters.

Cardy's crucial idea is that this partition function can be calculated
using the Hamiltonian with periodic boundary conditions $H^P$ which
generates translation in the $v$ direction.
By a conformal map
\begin{equation}
\xi = {\rm exp}(-{2\pi i\over \beta} w), \label{emap}
\end{equation}
we map the
cylinder with $H^P$
into the complex $\xi$ plane where the Virasoro generators $(L^P_n,
{\bar L}^P_n)$ are defined.  We
find that the Hamiltonian with periodic boundary conditions is given by
\begin{equation}
H^P={2\pi \over \beta}(L^P_0 + {\bar L}^P_0 -{c\over 12}). \label{ehamilP}
\end{equation}
It has both left and right movers.
Using (\ref{ehamilP}), the partition function is then
\begin{eqnarray}
Z^{(a,b)}_P({\tq})&=&\<< a|{\rm exp}(-lH^P)|b\>>\label{partmod} \\
&=& \tq^{-c/24} \<< a|\tq^{{1\over 2}(L^P_0 + {\bar L}^P_0)}|b\>> \\
{\rm where}~~ \tq&=&{\rm exp}(-{4\pi l \over \beta}).
\label{epartPab}
\end{eqnarray}
$|a\>>$ and $|b\>>$ are boundary states that obey the following
conditions.
On the cylinder, the boundary conditions at $v=0,l$ are
$J(w)={\bar J}({\bar w})$ and $T(w)={\bar T}({\bar w})$.
Using the conformal map (\ref{emap}), we get
\begin{eqnarray}
(J^P_n+{\bar J}^P_{-n})|b\>> &=& 0  \label{econds} \\
(L^P_n-{\bar L}^P_{-n})|b\>> &=& 0. \label{econdst}
\end{eqnarray}
We will see later that with the $U(1)$ Kac-Moody symmetry,
(\ref{econdst}) follows from (\ref{econds}).   Here, it is the
currents $(J,{\bar J})$ that classify the states into primaries, not
the Virasoro
generators.
Ishibashi \cite{Ishibashi} and Cardy \cite{Cardy89} showed that these
boundary states $|a\>>$ can be build out of linear combinations of the
Ishibashi states that by construction obey (\ref{econds}).
An Ishibashi state is given symbolically by
\begin{equation}
|j\>>=\sum_N |j;N\>> \otimes {\overline{|j;N\>>}} \label{eIshibashi}
\end{equation}
where $j$ denotes a primary state and $N$ denotes its $N^{th}$
descendent with normalization
$\<< j;N|j';N' \>> = \delta_{j,j'} \delta_{N,N'}$.
We will give the Ishibashi states explicitly in
our problem.
Then we can rewrite the partition function using (\ref{echardef}) and
(\ref{econdst}) as
\begin{equation}
Z_P^{(a,b)}({\tq})=\sum_j \<< a|j;0\>>\<< j;0|b\>> \chi_j(\tq). \label{epartP}
\end{equation}

We now equate the two partition functions (\ref{epartab}) and
(\ref{epartP}) to constrain the operator
content $n^i_{ab}$.  But note that one is a function of $q$ and the
other $\tq$.
We have to convert one into the other.
The difference in having an irrational conformal field theory versus a
rational one is that the sum in
(\ref{epartab}) and (\ref{epartP}) run over infinite number of primary
states in the irrational case.
Furthermore, for a rational conformal field theory, we can turn the
$q$ dependence in (\ref{epartab}) into $\tq$ by a linear
transformation given by the known modular $S$ matrix.  That is,
\begin{equation}
\chi_i(q) = \sum_j S_i^j \chi_j (\tq).
\label{eSmatrix}
\end{equation}
By equating the two partition
functions, one obtains
\begin{equation}
\sum_i n^i_{ab} S_i^j = \sum_j \<< a|j;0\>> \<<j;0|b\>>~.
\label{egnm}
\end{equation}
Here, with infinite
number of primaries, the transformation from $q$ to $\tq$ is not
given by (\ref{eSmatrix}) but by
\begin{equation}
\chi_Q(a,\delta;q)
={1 \over \sqrt{a}} \int dQ' e^{2\pi i(Q-{\delta\over 2\pi })Q'}
\chi_{Q'}({1\over a},0;\tq)
\end{equation}
where $\chi_Q$ is defined in (\ref{echaracter}).  This modular
transformation can be derived using the Gaussian integral and by
expressing $q={\rm exp}(-{\pi\beta \over l})$ and
$\tq={\rm exp}(-{4\pi l \over \beta})$.
Note that modular transformation requires the summation of a
continuous set of conformal towers $\chi_{Q'}$.
However, the modular transformation gives a discrete sum over a set of
conformal towers when we sum up the
contribution from each tower in the following way:
\begin{eqnarray}
\sum_{Q=-\infty}^\infty e^{i\delta_1 Q} \chi_Q(a,\delta_2;q)
=e^{({i\delta_1\delta_2 \over 2\pi})}{1\over \sqrt{a}}
\sum_{P=-\infty}^\infty e^{i\delta_2 P} \chi_{P}({1\over a},-\delta_1;\tq).
\label{emodtmn}
\end{eqnarray}
We derive this equation in the appendix.
Hence, we will express
the partition functions in terms of
\begin{equation}
\Omega (a,\delta_1,\delta_2;q) =
\sum_{Q=-\infty}^\infty e^{i\delta_1 Q} \chi_Q(a,\delta_2;q)
\label{eOmega}
\end{equation}
and the modular transformation (\ref{emodtmn})
becomes
\begin{equation}
\Omega (a,\delta_1,\delta_2;q) =
e^{({i\delta_1\delta_2 \over 2\pi})}{1\over \sqrt{a}}
\Omega ({1\over a},\delta_2,-\delta_1;\tq) ~.
\label{eOmodtmn}
\end{equation}
Without the $S$ matrix, we do not have (\ref{egnm}).  But we can still
equate the partition functions as Cardy did by using (\ref{eOmodtmn})
and solve for $n^i_{ab}$ of (\ref{epartab}) in the irrational case.
The solutions now rest on satisfying the following
consistency conditions on $n^i_{ab}$ of (\ref{epartab}) for each of
the infinite number of primaries $i$:
\begin{eqnarray}
\begin{array}{l}
n^i_{ab}~{\rm must~be~an~integer~for~any~pair}~(a,b)~{\rm and}
 \\
n^{i=0}_{aa}=1~{\rm for~each}~a~.
\end{array}
\label{econs}
\end{eqnarray}
The second consistency condition comes from demanding a unique vacuum
through the one to one correspondence between scaling dimensions of
operators and the finite size spectrum with identical boundary
conditions at both ends \cite{Cardy89}.

Recall that in the rational case, Cardy \cite{Cardy89,Cardy86} found the
boundary states
corresponding to spin up, down and free boundary conditions for the
Ising spins and gave
the finite spectrum for any pair of boundary conditions.
He showed that we can start from a
spectrum determined by a set of boundary conditions and obtain the other
spectra with other boundary conditions by the process of fusion.  The
case is similar in Kondo \cite{Afflecko,Affleckti}
and Heisenberg chain \cite{Affleckh}
where fusion with the
impurity spin give the finite size spectra with the new boundary
conditions.
In the present case where the $S$ modular transformation is
given by an integral, fusion does not seem to work.

We see that $J(w)={\bar J}({\bar w})$ and $T(w)={\bar T}({\bar w})$
are imposed at the boundary in Cardy's formalism and therefore exclude
the periodic boundary condition since no momentum or charge can
pass the boundary.  To incorporate the possibility of the periodic
boundary condition being one of the conformally invariant boundary fixed point,
we fold the system in half and double the bulk degrees of freedom.
In a finite size system with the two channels we have a set of
boundary conditions at both ends of the system.  The periodic boundary
condition in the unfolded system would correspond to having all the
momenta coming in through one channel go out the other.  This give the
perfect conductance case.  The open boundary condition will have all
the momenta coming in one channel reflected away in the same channel.
This give the zero conductance scenario.

More precisely, consider a finite size system extending from $-l$ to
$l$ where
the scattering potential or impurity is placed at the origin.  In
order to fold the system in half, we impose the same boundary
conditions at $-l$ and $l$ and identify the two points.  We expect
that the interaction at the
origin with the potential or impurity will renormalize into a boundary
condition at the origin.
Let us now see what the periodic and open boundary conditions on the $U(1)$
currents at the origin of the unfolded system become for the two
channel system.
Before folding as in
figure 1a,
we have for open boundary
$J(x=0_+,t)={\bar J}(x=0_+,t)$, $J(x=0_-,t)={\bar J}(x=0_-,t)$ and for
periodic boundary
$J(0_+,t)=J(0_-,t)$, ${\bar J}(0_+,t)={\bar J}(0_-,t)$ where $J$ and
${\bar J}$ are the left and right moving currents.
We folded the system about the origin in figure 1b so that the
currents in the two channels are related to the unfolded system by
\begin{eqnarray}
J(x>0)&=&J^1(x) ,~ {\bar J}(x>0)={\bar J}^1(x)  \nonumber \\
J(x<0)&=&{\bar J}^2(-x) ,~ {\bar J}(x<0)= J^2(-x).
\label{efold}
\end{eqnarray}
Therefore, periodic and open boundary conditions at $x=0$ in the
two channel system are
\begin{eqnarray}
J^1(0)&=&{\bar J}^1(0),~ J^2(0)={\bar J}^2(0)~~{\rm open}  \nonumber \\
J^1(0)&=&{\bar J}^2(0),~ J^2(0)={\bar J}^1(0)~~{\rm periodic}.
\label{ebcJ}
\end{eqnarray}

The finite size spectrum with appropriate boundary
conditions at $-l$, $l$ and $0$ is the same as the folded two channels
system.  Therefore, the folding process does not affect the calculation
of the partition function (\ref{epartab}).   With two channels, $H^P$ in
(\ref{ehamilP}) now becomes
\begin{equation}
H^P={2\pi \over \beta}(L^1_0 + {\bar L}^1_0 +L^2_0 + {\bar L}^2_0
-{c\over 6}). \label{ehamiltc}
\end{equation}
We have dropped the superscript $P$ in $L^P_0$ but understand that it
is distinguished from $L_0$ in (\ref{epartab}).
(\ref{econds}) becomes
\begin{equation}
( J^1_n+{\bar J}^1_{-n} + J^2_n+{\bar J}^2_{-n} )|a\>> = 0 . \label{econdstc}
\end{equation}
In the following sections, we will solve (\ref{econdstc}) and work out
the boundary states
corresponding to zero and perfect conductances
in both the
spinless and the spin-$1\over 2$ fermion cases.

\section{Spinless fermions}
In this section, we will illustrate our procedure in the spinless
fermion case before generalizing to the spinful case.    The plan is as
follows.  We will obtain the partition function (\ref{epartP}) by
constructing two boundary states that satisfy (\ref{econdstc}).  We label
the two boundary states by $|1\>>$ and $|2\>>$ but refer to them as
``open'' and ``periodic'' in anticipation that they correspond to zero and
perfect conductance.  We first need to obtain (\ref{ehamiltc}).  To do
that, we  bosonize the two channels of interacting fermions and obtain the
finite size spectrum for the bosons with periodic boundary conditions.
We then change basis from the two
channels to a more convenient even and odd basis.
In the even and odd basis, we give explicitly the open and periodic
boundary states.
We will obtain $Z^{(a,b)}_P(\tq)$
for the three combinations of pairs of the two boundary states.  By
modular transforming the three $Z^{(a,b)}_P(\tq)$, we obtain three
partition functions denoted by
${\cal Z}_{11}(q)$ ${\cal Z}_{22}(q)$ and ${\cal Z}_{12}(q)$.
To satisfy (\ref{econs}), we make an appropriate change in
normalization for the states $|1\>>$ and $|2\>>$.
In the appendix, we have, for these simple cases,
directly imposed the corresponding pairs of boundary conditions at the
two ends of the finite size system and obtained three
spectra. From these three spectra, we computed $Z_{pp}(q)$ $Z_{oo}(q)$
and $Z_{op}(q)$ as defined in
(\ref{epartab}) where $p$ and $o$ denotes periodic and open
boundary conditions.
In other cases where we do not have simple boundary conditions on the
fermions, we will have to rely on
using the boundary states and the consistency conditions (\ref{econs})
to compute the various partition functions.
Thus, we are able to verify with the appropriate normalizations that
$Z_{oo}(q)={\cal Z}_{11}(q)$, $Z_{pp}(q)={\cal Z}_{22}(q)$ and
$Z_{op}(q)={\cal Z}_{12}(q)$.
We will then give the equation for computing conductance in
terms of the boundary states.  By examining the operator content and
the ground state degeneracy, we find the conditions for the stability
of the two boundary fixed points.  This will be compared to
\cite{Kane}.  We will end the section by discussing the resonant
tunneling problem.

\subsection{Calculation of the partition functions}
The Hamiltonian can be considered to be as in Eq. (\ref{micHam}) with only a
nearest neighbor interaction term for the spinless fermions.
We are
interested in the low energy behavior and therefore we take the long
wavelength limit by expanding the fermion field about the Fermi momentum
$k_F$ \begin{equation}
\psi=e^{-ik_Fx}\psi_L + e^{ik_Fx}\psi_R.
\label{ekF}
\end{equation}
$\psi_L$ and $\psi_R$ are the low energy degrees of freedom.
In the continuum limit, the Hamiltonian is a relativistic theory
for $\psi_L$ and $\psi_R$ with a four fermi interaction\cite{AffleckLH}.
One can bosonize and parametrize the strength of the interaction by
a positive real number $R$.
More precisely,
we bosonize the left and right moving fermions by
\begin{equation}
\psi_L \sim \left[ {\rm exp}~-i({\phi \over 2R}+2\pi R \tphi) \right]
{}~~~~\psi_R \sim \left[ {\rm exp}~i({\phi \over 2R}-2\pi R
\tphi)\right]  \label{ebosonize}
\end{equation}
where $\phi=\phi_L + \phi_R$, $\tphi=\phi_L - \phi_R$
and $R={1\over
\sqrt{4 \pi}}$ at the free fermion point.
Here, we have already rescaled $\phi \rightarrow \phi /\sqrt{4\pi}R$
and $\tphi \rightarrow \sqrt{4\pi}R \tphi$ so that
the continuum Hamiltonian is
simply the one for a free boson, \begin{equation}
H={1 \over 2} \int_0^L [(\pa_x \phi)^2 + (\pa_t \phi)^2] dx.
\label{efree}
\end{equation}
We will often compare our results with Kane and Fisher's and $R$ is
related to their interaction strength $g$ \cite{Kane} by
$4\pi R^2=1/g$ in the spinless case.

To avoid confusion when we compute $Z^{(a,b)}_P(\tq)$, we will take
the space
to be periodic in $0 \leq x \leq L$ and $t$ to be the Minkowski time.
At the end we will substitute $L=\beta$, the inverse temperature.
Let us concentrate on one channel for the moment.
To obtain $H^P$, we need to specify the boundary conditions on the low
energy fields.
In computing $Z_{ab}(q)$ of (\ref{epartab}) at a finite temperature,
we impose antiperiodic
boundary condition on the fermions $\psi_{L,R}$ or equivalently
periodic boundary condition on
the bosons $\phi$ and $\tphi$ in the imaginary {\it time} direction.
Therefore, when we compute $Z^{(a,b)}_P(\tq)$ now
where we have interchanged the role of space and imaginary time, we
impose
(anti)periodic boundary condition on the (fermions) bosons in the
 {\it space} direction.
By imposing antiperiodic boundary condition,
$\psi_{L,R}(x)=-\psi_{L,R}(x+L)$, we use
(\ref{ebosonize}) to obtain the boundary
conditions on $\phi$ and $\tphi$
\footnote{One need to use the fact $[\phi^L,\phi^R]=i/4$.}
\begin{eqnarray}
Q = \phi(L)-\phi(0)=2\pi nR ~~~~
\Pi &=& \tphi(L)-\tphi(0)= {m \over 2R} \label{ebcboson} \\
{\rm where}~~n&=&m~~(\mod \ 2). \label{egluing}
\end{eqnarray}
We refer to the restriction on the quantum numbers in (\ref{egluing})
as the ``gluing conditions''.
We see that $\phi$ and $\tphi$ are bosons compactified on circles with
radii $R$ and $1/4\pi R$ respectively.
With these boundary conditions, we can write down a mode expansion for
the boson $\phi$ following \cite{Eggert}
\begin{eqnarray}
\phi(x,t)&=&\phi_0 + {1\over 2}(\Ph + \Qh)\tpL +{1\over 2}(\Ph -
\Qh)\tmL  \nonumber \\
&+& \sum_{n=1}^{\infty}{1 \over \sqrt{4\pi n}}
[e^{-2\pi i n \tpL}a^L_n + e^{-2\pi i n \tmL}a^R_n +h.c.]
\label{emode}
\end{eqnarray}
where the eigenvalues of $\Qh$ and $\Ph$ are $Q$ and
$\Pi$ defined in (\ref{ebcboson}).
$\tphi (x,t)$ has a similar mode expansion
\begin{eqnarray}
\tphi(x,t)&=&\tphi_0 + {1\over 2}(\Ph + \Qh)\tpL -{1\over 2}(\Ph -
\Qh)\tmL  \nonumber \\
&+& \sum_{n=1}^{\infty}{1 \over \sqrt{4\pi n}}
[e^{-2\pi i n \tpL}a^L_n - e^{-2\pi i n \tmL}a^R_n +h.c.].
\label{emodet}
\end{eqnarray}
The nonzero canonical commutation relations are
$[a_n^{L},a_m^{L\dagger}]=[a_n^{R},a_m^{R\dagger}] =\delta_{nm}$ and
$[\Qh,\tphi_0]=[\Ph, \phi_0]=-i$.
Substituting the mode expansion
(\ref{emode}) into (\ref{efree}), the Hamiltonian becomes
\begin{equation}
H={2\pi \over L}[{1\over 4\pi}(\Ph^2 + \Qh^2) +
\sum_{n=1}^\infty n({\hat m}^L_n + {\hat m}^R_n) - {1\over 12}]
\label{eHmode}
\end{equation}
where ${\hat m}^L_n=a^{L \dagger}_na^L_n$ and ${\hat m}^R_n=a^{R
\dagger}_na^R_n$
are the boson occupation number operators and the ${1\over 12}$ term is the
ground state energy from (\ref{ehamilP}) with $c=1$ for the boson
theory.

The conserved Virasoro and $U(1)$ Kac-Moody currents are given by
\begin{eqnarray}
T&=&:(\pa_+ \phi)^2:, ~~{\bar T}=:(\pa_- \phi)^2:
\nonumber \\
J&=&{1\over 2\pi R}\pa_+ \phi,~~{\bar J}=-{1\over 2\pi R}\pa_- \phi
\label{ecurrent}
\end{eqnarray}
where $\pa_{\pm}={1\over 2}(\pa_t \pm \pa_x)$ and $:~:$ denotes normal
ordering.
To get the condition (\ref{econdstc}), we substitute (\ref{emode})
into (\ref{ecurrent}) and expand the current in terms of the boson
operators at $t=0$ by
$$J_n= \int_0^L dx~ {\rm exp}({2\pi i n x \over
L})J(x,0)~~
{\rm and}~~ {\bar J}_n= \int_0^L dx~ {\rm exp}(-{2\pi i n x \over
L}){\bar J}(x,0).$$
We obtain for $n > 0$
\begin{eqnarray}
J_0&=&{1 \over 4\pi R}(\Ph + \Qh),~~J_{n}=-\sqrt{n\over 4\pi
R^2}ia^L_n,~~J_{-n}=J^\dagger_n
\nonumber  \\
{\bar J}_0&=&{1 \over 4\pi R}(-\Ph + \Qh),~~{\bar J}_{n}=\sqrt{n\over 4\pi
R^2}ia^R_n,~~{\bar J}_{-n}={\bar J}^\dagger_n.
\label{eJmode}
\end{eqnarray}

A primary state with respect to the U(1) Kac-Moody algebra is
\begin{equation}
|Q,\Pi\>> =e^{i(Q\tphi_0 + \Pi \phi_0)} |0\>> \label{eprimary}
\end{equation}
since we can see from (\ref{eJmode}) that
\begin{eqnarray}
J_{n>0}|Q,\Pi\>>&=&0,~~~~
{\bar J}_{n>0}|Q,\Pi\>>=0,  \nonumber \\
J_{0}|Q,\Pi\>>&=&{1 \over 4\pi R}(\Pi + Q)|Q,\Pi\>>,~~{\rm and }~~
{\bar J_{0}}|Q,\Pi\>>={1 \over 4\pi R}(-\Pi + Q)|Q,\Pi\>>. \nonumber
\end{eqnarray}
For $Q$ or $\Pi$ nonzero, this state is also a primary state with
respect to the Virasoro algebra $(L_n,{\bar L_n})$ where
\begin{equation}
L_n={2\pi R^2} \sum_{m=-\infty}^\infty :J_mJ_{n-m}:~.
\end{equation}
Since $L_n$ are bilinear in $J_n$, in the presence of the $U(1)$
symmetry we use the more fundamental $J_n$ to classify the states
into primaries and descendents.
The descendents of the primary state (\ref{eprimary}) are given by
\begin{equation}
e^{i(Q\tphi_0 + \Pi \phi_0)}\prod_{n=1}^\infty
{(a^{L \dagger}_n)^{m^L_n} \over ({m^L_n}!)^{1\over 2}}
{(a^{R \dagger}_n)^{m^R_n} \over ({m^R_n}!)^{1\over 2}} |0\>>
\end{equation}
where $m^{L,R}_n$ are the occupation numbers.
An Ishibashi state is one that by construction  satisfies
(\ref{econds}).  By using (\ref{eJmode}), we can show that
\begin{eqnarray}
|Q,\Pi\>>_I &=& e^{i \Pi \phi_0}
\prod_{n=1}^\infty \sum_{m_n=0}^\infty
{{(-a^{L \dagger}_n a^{R \dagger}_n)}^{m_n} \over {m_n}!} |0\>>
\nonumber \\
&=& e^{i \Pi \phi_0}\prod_{n=1}^\infty e^{-a^{L \dagger}_n a^{R
\dagger}_n} |0\>> \label{eIstate}
\end{eqnarray}
is the desired Ishibashi state.
We can put the above Ishibashi state into the form (\ref{eIshibashi})
by rewriting the product of sums as sum of products as follows,
$$\prod_{n=1}^\infty \sum_{m_n=0}^\infty
=\sum_{N=0}^\infty \sum_{\{m_n\}}{^\prime}~ \prod_{n=1}^\infty$$
where the prime in the sum denotes the restriction
$\sum_n m_n =N$.
A boundary state is a linear combinations of such Ishibashi states.
The open and periodic boundary states exist only in
the two channel system which we will turn to next.

For a system with two channels, we work with two copies of bosons
$\phi^1$ and $\phi^2$.  We will see that it is advantageous to project
the two channels into an even and odd basis similar to the two
impurity Kondo problem \cite{Affleckti}.  In the interacting fermion
picture, we have a problem since this will generate nonlocal interactions.
However, we can do so now after bosonization since we have a free
theory.
To define and see the advantage of the even and odd basis, we go back to
the current conservation for the two channel system
\begin{equation}
J^1+J^2 -{\bar J}^1-{\bar J}^2=0, ~~~~ T^1+T^2 -{\bar T}^1-{\bar T}^2=0
\end{equation}
and substitute (\ref{ecurrent}).  By defining
\begin{equation}
\phi^{e,o}={1\over \sqrt 2}(\phi^1 \pm \phi^2), \label{ebeo}
\end{equation}
we get
\begin{eqnarray}
J^e -{\bar J}^e&=&0, ~~~~ T^e+T^o -{\bar T}^e-{\bar T}^o=0 \nonumber \\
{\rm where}~~J^e=J^1+J^2 &=& {1\over \sqrt{2}\pi R}\pa_+ \phi^e,~~
T^e=:(\pa_+ \phi^e)^2:~~{\rm and}~~
T^o=:(\pa_+ \phi^o)^2:~~.
\label{econdseo}
\end{eqnarray}
$J^e$ is the total current of the two channels.
Notice that $J^o=J^1-J^2$ is not present in the constraints.
By combining the two current equations in (\ref{econdseo}), we deduce
that $T^o={\bar T}^o$.  That is, in the even channel, the boundary
preserves the
$U(1)$ Kac-Moody symmetry but in the odd channel, the boundary only
preserves the smaller conformal symmetry.
It is important to note that in the bulk, however, both $J_e$ and
$J_o$ are conserved currents.

To get the Ishibashi states in the even and odd basis, we need to
expand $\phi^{e,o}$ in modes.  We start with the mode expansions of
$\phi^1$ and $\phi^2$ as in (\ref{emode}), noting that there are
gluing conditions (\ref{egluing})
between $Q^1$ and $\Pi^1$ and
similarly  $Q^2$ and $\Pi^2$.
By (\ref{ebeo}), we obtain mode expansion for $\phi^{e,o}$ as in
(\ref{emode}) with
\begin{eqnarray}
a_n &\rightarrow & a^{e,o}_n = {1\over \sqrt{2}}(a^1_n \pm a^2_n)~~~~
{\rm and} \nonumber \\
\Pi &\rightarrow  & \Pi^{e,o}={1\over \sqrt{2}}(\Pi^1 \pm \Pi^2)
\equiv {1\over \sqrt{2}}{m^{e,o}\over 2R} \nonumber \\
Q &\rightarrow & Q^{e,o}={1\over \sqrt{2}}(Q^1 \pm Q^2)
\equiv  \sqrt{2}\pi R n^{e,o}
\label{etmneo}
\end{eqnarray}
where $m^{e,o}= m^1+m^2$ and $n^{e,o}= n^1+n^2$.
The gluing conditions between $m^{e,o}$ and $n^{e,o}$ can be derived
from $m^{1,2}=n^{1,2} ({\rm mod}~2)$.  We obtain
\begin{equation}
m^{e,o}=n^{e,o}~~({\rm mod}~2)~~~~ {\rm and} ~~~~ m^e+m^o+n^e+n^o=0 ~~
({\rm mod}~4)
\label{egluingeo}
\end{equation}
Note that the even and odd channels are not decoupled.

The Ishibashi states for the even channel must be (\ref{eIstate})
\begin{equation}
|n^e=0,m^e\>>^e_I \equiv |Q^e=0,\Pi^e\>>^e_I= e^{i \Pi^e
\phi^e_0}\prod_{n=1}^\infty e^{-a^{eL \dagger}_n a^{eR \dagger}_n}
|0\>>^e.
\label{eIe}
\end{equation}
However, in the odd channel, we do not necessarily have such
Ishibashi states because the odd $U(1)$ current is not constrained.
But for the open and periodic boundary conditions, something special
happens.  Transforming the boundary conditions (\ref{ebcJ}) into even
and odd sector using (\ref{econdseo}), we arrive at
\begin{eqnarray}
J^e(0)-{\bar J}^e(0)&=&0,~ J^o(0)-{\bar J}^o(0)=0~~{\rm open} \\
J^e(0)-{\bar J}^e(0)&=&0,~ J^o(0)+{\bar J}^o(0)=0~~{\rm periodic}.
\end{eqnarray}
Therefore, we further have $U(1)$ conservation at the boundary in the
odd channel for open boundary and maximal violation of the odd $U(1)$
charge at the boundary for periodic boundary.
The periodic odd channel here resembles the problem of monopole-catalyzed
baryon decay where the baryon number conservation is maximally
violated at the boundary. \cite{Sagi}
Using the conformal map (\ref{emap}),
we require that the open Ishibashi state in the odd channel
also be annihilated by $J^o_n+J^o_{-n}$.  We then use the same
Ishibashi state as (\ref{eIe}) with $e \rightarrow o$.
Gluing this odd Ishibashi state with the even one leads to
the open Ishibashi state.
For the periodic Ishibashi state in the odd channel,  we impose that it
be annihilated by $J^o_n-J^o_{-n}$, giving us
\begin{equation}
|n^o,m^o=0\>>^o_I \equiv |Q^o,\Pi^o=0\>>^o_I= e^{i Q^o
\tphi^o_0}\prod_{n=1}^\infty e^{a^{oL \dagger}_n a^{oR \dagger}_n}
|0\>>^o.
\label{eIo}
\end{equation}
When glued  with the even channel, this leads to the periodic
Ishibashi state.

Since we have an infinite number of primaries, we will have a
sum over infinite number of Ishibashi states to obtain a boundary
state.
Consider the following two boundary states, each a linear combinations
of the Ishibashi states,
\begin{eqnarray}
|1\>> &=& \sum_{m^e,m^o}{^\prime}~ C_{m^e,m^o} |0,m^e\>>_I^e \otimes
|0,m^o\>>_I^o
\label{ebstateso}\\
|2\>> &=& \sum_{m^e,n^o}{^\prime}~ C_{m^e,n^o} |0,m^e\>>_I^e \otimes
|n^o,0\>>_I^o
\label{ebstatesp}
\end{eqnarray}
where the primes denote summing over the quantum numbers allowed by the
gluing conditions (\ref{egluingeo}).
The boundary states $|1\>>$ and $|2\>>$ determine
$Z^{(1,1)}_P(\tq)$,$Z^{(2,2)}_P(\tq)$ and $Z^{(1,2)}_P(\tq)$,
which in turn give
${\cal Z}_{11}(q)$, ${\cal Z}_{22}(q)$ and ${\cal Z}_{12}(q)$ by a
modular transformation.
By imposing the consistency condition (\ref{econs}) on the partition
functions ${\cal Z}$'s,
we find that $C_{m^e,m^o}$ and $C_{m^e,n^o}$ can at most be phases
${\rm exp}(im^e \alpha + im^o \beta)$ and
${\rm exp}(im^e \alpha ' + in^o \beta ')$.
These phases amount to the chemical potentials in the two bulk
channels and do not affect the boundary physics.
We will proceed with $C_{m^e,m^o}=C_{m^e,n^o}=1$ to illustrate the
calculation.

For two channels, $H^P$ is given by (\ref{ehamilP}) and expanded in
modes, each channel is given by (\ref{eHmode}).
In terms of even and odd basis, we can rewrite $H^P$ using
(\ref{etmneo}) as
\begin{equation}
H^P={2\pi \over L}[{1\over 4\pi}(\Ph^{e2} + \Qh^{e2}+ \Ph^{o2} + \Qh^{o2}) +
\sum_{n=1}^\infty n({\hat m}^{eL}_n + {\hat m}^{eR}_n + {\hat m}^{oL}_n
+ {\hat m}^{oR}_n) - {1\over 6}]
\label{eHPeo}
\end{equation}
Using (\ref{epartPab}) generalized to the two channel problem and
(\ref{eHPeo}) with $L=\beta$, we obtain for the boundary state
(\ref{ebstateso})
\begin{eqnarray}
Z^{(1,1)}_P(\tq)&=&\<< 1|e^{-lH^P}|1\>> \nonumber \\
&=&\tq^{-{1\over 12}}\sum_{m^e,m^o}{^\prime}~
\sum_{m^{e,o}_1,m^{e,o}_2,\ldots=0}^\infty
\tq^{[{1\over 64\pi R^2}(m^{e2}+ m^{o2})+ \sum_{n=1}^\infty n(m^e_n + m^o_n)]}
\end{eqnarray}
where $\tq={\rm exp}(-{4\pi l \over \beta})$.  Once again, the prime in the
sum denotes summing over quantum numbers allowed by the gluing
conditions \ref{egluingeo}.  The Ishibashi states sets $Q^e$ and $Q^o$
to zero and $m^{eL}_n = m^{eR}_n =m^e_n$ and $m^{oL}_n = m^{oR}_n=m^o_n$ in
(\ref{eHPeo}).  Solving the gluing conditions (\ref{egluingeo}) with
$n^e=n^o=0$, we see that $m^{e,o}=2k^{e,o}$ are even where $k^{e,o} \in
{\bf I}$ and $k^e +k^o=0 \pmod 2$.  Substituting into (\ref{eHPeo}), we
get
\begin{equation}
Z^{(1,1)}_P(\tq)={1\over \eta (\tq)^2}\sum_{k^e+k^o=0 \pmod 2}
\tq^{[{1\over 16\pi R^2}(k^{e2}+k^{o2})]}
\end{equation}
where the Dedekind function $\eta$ is obtained by summing over $m^e_n$
and $m^o_n$.
Solving the constraints by letting $k^{e,o}=k\pm l$, $k,l \in {\bf I}$,
we finally obtain in terms of $\Omega$ defined in (\ref{eOmega})
\begin{eqnarray}
Z^{(1,1)}_P(\tq)&=& \Omega ({1\over 4\pi R^2},0,0;\tq)^2 \nonumber \\
&=&   4\pi R^2 \Omega ( 4\pi R^2,0,0;q)^2
\equiv {\cal Z}_{(1,1)}(q).
\label{ezoo} \end{eqnarray}
Similarly, we obtain
\begin{equation}
Z^{(2,2)}_P(\tq)={\cal Z}_{(2,2)}(q)=  \Omega ({1\over 2\pi R^2},0,0;q)
\Omega (8\pi R^2,0,0;q)+
\Omega ({1\over 2\pi R^2},0,\pi;q) \Omega (8\pi R^2,0,\pi;q)
\label{ezpp}
\end{equation}
This partition function is modular invariant, that is,
$Z^{(2,2)}_P(\tq)=Z^{(2,2)}_P(q)$.
The reason is that we have periodic boundary conditions for the bosons
in both the
time and space directions, or equivalently, antiperiodic boundary
conditions on the fermions $\psi_{L,R}$ in both directions.
It needs some explanation to compute ${\cal Z}_{(1,2)}$.  The mixed
matrix element between the boundary states (\ref{ebstateso}) and
(\ref{ebstatesp}) sets $m^o=n^o=0$ giving
\begin{equation}
Z^{(1,2)}_P(\tq)=
\tq^{-{1\over 12}}\sum_{m^e}{^\prime}~
\sum_{m^{e,o}_1,m^{e,o}_2,\ldots=0}^\infty
(-1)^{m^o_n} ~
\tq^{[{m^{e2}\over 64\pi R^2} + \sum_{n=1}^\infty n(m^e_n +m^o_n)]}.
\end{equation}
The gluing condition sets $m^e=4k$, $k \in {\bf I}$.  We then get
\begin{eqnarray}
Z^{(1,2)}_P(\tq)&=& \Omega ({1\over 2\pi R^2},0,0;\tq)~\tq^{-{1\over 24}}
\sum_{m^{o}_1,m^{o}_2,\ldots=0}^\infty \prod_{n=1}^\infty (-\tq)^{m^o_n}
\nonumber \\
&=& \Omega ({1\over 2\pi R^2},0,0;\tq)~ W(\tq) \nonumber   \\
{\rm where}~~ W(\tq)&=& \tq^{-{1\over 24}}\prod_{n=1}^\infty{1\over
1+\tq^n}
= \Omega(2,\pi,0;\tq).
\end{eqnarray}
The last equality in $W(\tq)$ is given by the Jacobi triple product
identity \cite{Yang}. (See appendix for a similar derivation).
By modular transforming, we get
\begin{eqnarray}
{\cal Z}_{(1,2)}(q)&=& \sqrt{4\pi R^2} \Omega (2\pi R^2,0,0;q)
W_+(q) \nonumber \\
{\rm where}~~ W_+(q)&=& {1\over 2}\Omega({1\over 2},0,\pi;q).
\label{ewp}
\end{eqnarray}

As argued before, we must be allowed to impose any pairs of valid
boundary conditions to the bulk.  The criterion is that
for any pairs of boundary states, the partition
functions ${\cal Z}_{(a,b)}$ generated must satisfy (\ref{econs}).
We see that if we normalize the state $|1\>>$ by $1 /4\pi R^2$, then
all the partition functions have unit integer coefficients.
Comparing these partition functions with $Z_{pp}(q)$ $Z_{oo}(q)$ and
$Z_{op}(q)$ worked out in the
appendix from imposing
the boundary conditions directly, we
conclude that
\begin{equation}
|{\rm periodic}\>> = |2\>> ~~{\rm and}~~ |{\rm open}\>> = {1\over
\sqrt{4\pi R^2}}|1\>>
\label{enbstates}
\end{equation}
are the appropriate boundary states.

\subsection{Conductance}
We define the charge conductance beginning with the Kubo formula as in
\cite{Kane},
\begin{equation}
G=\lim_{\omega \rightarrow 0} {4\pi^2 e^2 \over \hbar (2l)^2 \omega}
\int_{-l}^ldx \int_{-l}^ldy \int_{-\infty}^{\infty}d\tau e^{i\omega \tau}
\<< j(x,\tau) j(y,0)\>>
\label{econduct}
\end{equation}
where $j(x,\tau)=J-{\bar J}$ is the spatial component of the current
in the unfolded system.
Folding into a two channel system by (\ref{efold}) and going into the
even and odd basis by $J^{e,o}=J^1 \pm J^2$ (\ref{econdseo}), we see that
$\int_{-l}^l dx~j(x,\tau)=\int_0^l dx (J^o-{\bar J}^o)$.

To evaluate the conductance, we use the following correlation
functions:
\begin{eqnarray}
\<<J^o(x,\tau)J^o(y,0)\>> &=& {1\over 2\pi R^2}~{1\over 4\pi^2[\tau +i(x-y)]^2}
\nonumber \\
\<<{\bar J}^o(x,\tau){\bar J}^o(y,0)\>> &=& {1\over 2\pi R^2}~{1\over
4\pi^2 [\tau -i(x-y)]^2} \nonumber \\
\<< J^o(x,\tau){\bar J}^o(y,0)\>> &=& {A\over
4\pi^2 [\tau +i(x+y)]^2} \nonumber \\
\<<{\bar J}^o(x,\tau) J^o(y,0)\>> &=& {A\over
4\pi^2 [\tau -i(x+y)]^2}~. \nonumber
\end{eqnarray}
$A$ is sensitive to the boundary condition and is given by
\cite{Cardy91}\cite{AffleckG}
$$A=-{\<< 0|J^o_{n=1} {\bar J}^o_{n=1} | B\>> \over \<< 0 | B\>>},$$
where $|B\>> $ is the boundary state
and $J^{o\dagger}_{n=1} {\bar J}^{o\dagger}_{n=1} |0\>>$ is the first
descendent state of the vacuum with respect to the $U(1)$ algebra.  We
use only the first descendent
because the $U(1)$ charge operator is the first descendent of the
identity operator in the $U(1)$ theory and it is a primary operator
with respect to the Virasoro algebra.
After doing a contour integral in the complex $\tau$ plane,
taking the zero frequency limit and then performing the spatial
integrals,  we find
\begin{equation}
G={(1-2\pi R^2A) \over 2}{1\over 4\pi R^2}{e^2\over h}
\end{equation}
For $|B\>> =|{\rm open}\>>$, we see from (\ref{ebstateso}) that
$A=1/2\pi R^2$,
therefore
the conductance vanishes.
For $|B\>> =|{\rm periodic}\>>$, $A=-1/2\pi R^2$
and we obtain
\begin{equation}
G={1\over 4\pi R^2}{e^2\over h}
\end{equation}
which reproduces what Kane and Fisher got when we use $4\pi R^2=g^{-1}$.

\subsection{Operator content and ground state degeneracies}
We can find out the operator content (the dimensions of the boundary
operators) from the finite
size spectrum or the partition function when identical boundary
conditions are applied at both ends \cite{CardyLH,Cardy86}.
The partition functions equivalent to the ones given in
the appendix are
\begin{eqnarray}
Z_{(o,o)}(q)&=&[{1\over \eta (q)}\sum_{k}
q^{{2\pi R^2}k^2}]^2
\label{epartoo}\\
Z_{(p,p)}(q)&=&{1\over \eta (q)^2}\sum_{m+n=0 \pmod 2}
q^{{m^2\over 16\pi R^2}+n^2\pi R^2}
\label{epartpp}
\end{eqnarray}

Let's consider the open-open case first.  We have two decoupled
channels indicated by the complete square in $Z_{(o,o)}(q)$ and
therefore double the boundary operators of the one channel case.
Due to the decoupling of the two channels, it only make sense that we
think of the channels as $1$ and $2$ but not even and odd.
In the one channel case, we can read off from (\ref{epartoo}) that the
lowest surface dimensions of the primary operators are $0~
(k=0)$,$2\pi R^2~(k=\pm 1)$,
$8\pi R^2~(k=\pm 2)$, etc..
We will think of this problem in the purely left moving
formalism as indicated in the introduction. Recall that the dimension of an
operator $e^{i\alpha \phi_L}$ in a purely left moving system is
${\alpha^2 \over 8\pi}$ \cite{Ginsparg}.  From this
we can write down the corresponding operators with the above
dimensions, namely the identity operator, $e^{\pm 4\pi i R\phi_L}$,
$e^{\pm 8\pi i R\phi_L}$, etc.. This is the operator content for the
open boundary fixed point.

To analyze the stability of the boundary fixed point, we bring the two
channels together and make a local perturbation
about the open boundary.
We couple two of the above boundary operators, one from each channel.
These coupled operators are allowed perturbations only if they have the
symmetries of the Hamiltonian, in particular, they must be real and
$U(1)$ invariant. The $U(1)$ transformation is $\psi \rightarrow
e^{-i\alpha} \psi$.  Using equations (\ref{ekF}) and
(\ref{ebosonize}), we see that to effect this transformation by the
bose fields, we need $\phi \rightarrow \phi$ and $\tphi \rightarrow
\tphi + {\alpha \over 2\pi R}$.
Using $\phi=\phi_L + \phi_R$ and $\tphi=\phi_L
-\phi_R$, we find
\begin{equation}
\phi_L \rightarrow \phi_L + {\alpha \over 4\pi R}
{}~{\rm and}~~
\phi_R \rightarrow \phi_R - {\alpha \over 4\pi R}.
\label{ebLtmn}
\end{equation}
The lowest dimensional coupled
operators allowed by the $U(1)$ symmetry are
${\cal O}=Re[e^{ 4\pi i R(\phi^{1L}-\phi^{2L})}]$ or $i\{Im[e^{ 4\pi i
R(\phi^{1L}-\phi^{2L})}]\}$
which have
dimensions $4\pi R^2$.
The latter operator would be eliminated if we impose parity symmetry.
The operator ${\cal O}$ enters the Hamiltonian
as  $H_{int}=\lambda\int \delta (x) {\cal O} dxdt$, where $\lambda$ is
the coupling.  Counting the dimension of
$\delta (x)$ as one, we see that this is a relevant perturbation about the
open boundary when ${\rm dim}({\cal O}) =4\pi R^2 <1$. This corresponds to
the term $\left[q^{2\pi R^2}\right]$ in Eq. (\ref{epartoo}). In other words,
the open boundary  condition is stable as long as the bulk interaction is
repulsive,  $4\pi R^2 >1$.

Let us turn to the periodic case.  Here, both left and right movers
are present and the dimensions of
the operators are the sum of the left and the right's.
We will think of this system as having one channel of left and right
movers on length $2l$.
Furthermore, only $U(1)$ invariant operators are allowed in perturbing
the Hamiltonian.    As described above, $\phi$ is $U(1)$
invariant.  Any function of $\tphi$ will not be $U(1)$ invariant and
therefore not allowed to enter $H_{int}$.
Noting that in (\ref{epartpp}), dimensions ${m^2 \over 16\pi R^2}$
correspond to operators $e^{\pm im \phi /2R}$ and $n^2\pi R^2$
correspond to  $e^{\pm 2\pi inR\tphi}$, we set $n=0$.  The lowest
dimension allowed operators from $Z_{p,p}$ are the identity,
$e^{\pm i \phi / R}$ and $e^{\pm  2i \phi /R}$ with
dimensions $0$, ${1\over 4\pi R^2}$ and ${1\over \pi R^2}$.
If we perturb locally in the periodic system, the most important
operators $e^{\pm i \phi /R}$ will appear in $H_{int}$.
Parity invariance does not have much of an effect here:
it only restricts from the two possible linear combinations of $e^{\pm i
\phi / R}$ entering the Hamiltonian to one, $cos(\phi /R)$.
This operator
is relevant when ${1\over 4\pi R^2}<1$ and therefore the periodic
boundary condition is stable if ${1\over 4\pi R^2}>1$.  We can patch
this result nicely with the above open-open one and give precisely
what is in \cite{Kane}.

We have one more tool to decide stability of the boundary fixed
points.  It is the ground state degeneracy theorem of
\cite{Affleckgsd}.  The ground state degeneracy is the
universal number appearing
in front of the partition function in the limit ${l\over \beta}
\rightarrow \infty$.  It is not necessarily an integer because we have
an infinite length system.
It is proposed that under renormalization of
boundary interactions, the ground state degeneracy decreases.
Equivalently, we can associate the ground state degeneracy as the
normalization factor for the boundary states (\ref{enbstates}).
In our problem, we have ground state degeneracy for the periodic case
$g_p=1$ and $g_o = {1\over \sqrt{4\pi R^2}}$ for the open boundary.
The g-theorem nicely reproduce the above results:
$g_p<g_o$ is the
stability condition for the periodic boundary condition  and vice
versa.

\subsection{Resonant tunneling}
In the resonant tunneling case, there is just a minor adjustment to
the above reasoning.  There is now an extra impurity degree of
freedom which has zero scaling dimension as far as renormalization goes.
About open boundaries,  the impurity couples to the lowest
dimension operators $e^{\pm 4\pi i R\phi^L}$ via hopping:
e.g. $(t_0 \psi_1^\dagger \eta_I + \hbox{h.c.})$, where $\eta_I$ is the
impurity fermion field, corresponding to $\psi_0$ in Eq. (\ref{micHam}).
Hence the dimension of this perturbation interaction is $2\pi R^2$.
About the periodic case with parity invariance, Kane and Fisher
\cite{Kane} showed that
for resonant tunneling to take place, one
has to adjust the chemical
potential at the impurity site $\mu_0$ so that the probabilities of it being
empty and occupied are equal.  This fine tuning of the potential to
achieve resonance is equivalent to fine tuning away the lowest dimension
operator $cos(\phi /R)$ in the periodic system.
In the case without parity invariance, one has to adjust two
parameters to eliminate $e^{\pm i \phi / R}$ and achieve resonance.
The next
lowest dimension operator
is  $e^{\pm  2i \phi / R}$ of dimension ${1\over \pi R^2}$.
We see that the results about the two boundary fixed points do not
match nicely as before.
The open fixed point is stable when $2\pi R^2>1$ but the periodic
fixed point is stable when $\pi R^2<1$ as in \cite{Kane}.
Taking a hint from \cite{Kane}, we see that
about the open fixed point, the descendent of the identity operator
$J$ can also couple with the impurity and enter $H_{int}$ through
an induced interaction
$K~\psi^\dagger \psi \eta_I^\dagger \eta_I \sim K\pa_x\phi~\eta_I^\dagger
\eta_I$.  The descendents usually have
higher dimensions than the primaries and therefore not important.
Here, the dimension of the interaction is one and therefore a marginal
coupling of the descendent $J$ with the impurity. This together with
the hopping terms generate a Kosterlitz-Thouless type renormalization
flow on the $t-K$ plane which was used to
explain the disparities between the different
stability of the two boundary fixed points \cite{Kane}.
We agreed with Kane and Fisher's analysis.

Let us see what the ground state degeneracy says about the resonant
case.  For the open boundary at resonance, $g_r=2g_o$ where the
factor of two is due to the two states (empty or filled) at the decoupled
impurity site.  Therefore, $g_r={1\over \sqrt{\pi R^2}}$.  For the
periodic boundary condition, the impurity is absorbed by the continuum.
Thus $g_p=1$.  The stability of the periodic fixed point is given
$g_p<g_r$ which agrees with the above results.
There appear to be a discrepancy between the predictions for the stability of
the open boundary from the g-theorem ($g_r<g_p$)
and the operator content ($R^2> {1\over 2\pi}$).
But we see from
 the Kosterlitz-Thouless type renormalization
flow on the $t-K$ plane\cite{Kane} that for a given initial value of $K$ for
$R^2> {1\over 2\pi}$, there
is a critical bare $t$ that flows into the open fixed point
($t=0$) and then to the periodic fixed point ($t \rightarrow \infty$).
This renormalization flow holds until $R^2< {1\over \pi}$ beyond which
$t=0$ is a stable fixed point.
This flow is consistent with the $g$-theorem for $R^2< {1\over
\pi}$, where it occurs.

\section{spin-${1\over 2}$ fermions}
We will reproduce the same steps for the more complicated spinful
case.  One may begin with the lattice fermion model (\ref{micHam}) and
derive the low energy continuum free boson Luttinger liquid as in
\cite{Shankar}.  We will only give the bosonization rules.

\subsection{calculation of partition functions}
We concentrate on one channel for the moment.
We first take the low energy long wavelength
limit by expanding the fermion field about the Fermi momentum $k_F$
\begin{equation}
\psi_\alpha=e^{-ik_Fx}{\psi_L}_\alpha(x) +
e^{ik_Fx}{\psi_R}_\alpha(x) \label{ekFs}
\end{equation}
where $\alpha = \uparrow$ or $\downarrow$.
{}From the interacting fermions in the bulk, we obtain a free
theory of bosons with charge and spin interactions parametrized by the
radii $R_c$ and $R_s$.
That is, we bosonize the low energies left and right moving fermions by
\begin{eqnarray}
\psi_{L \uparrow ,\downarrow } \sim \left[
{\rm exp}~-i({\phi_c \over 2R_c}+\pi R_c \tphi_c \pm
{\phi_s \over 2R_s} \pm \pi R_s \tphi_s)\right]
\nonumber \\
\psi_{R \uparrow, \downarrow} \sim \left[
{\rm exp}~i({\phi_c \over 2R_c}-\pi R_c \tphi_c \pm
{\phi_s \over 2R_s} \mp \pi R_s \tphi_s) \right]~. \label{ebosonizes}
\end{eqnarray}
Here $\phi_{c,s}$ are linear combinations of the bosons, $\phi_{\uparrow
,\downarrow}$ introduced to represent the two fermion fields
$\psi_{\uparrow,\downarrow}$:
\begin{eqnarray}
\phi_c &\equiv &(\phi_\uparrow +\phi_\downarrow )/\sqrt{2}\nonumber \\
\phi_s &\equiv &(\phi_\uparrow -\phi_\downarrow )/\sqrt{2}
\end{eqnarray}
They represent the charge and spin degrees of freedom respectively.
At the free
fermion point, $R_c = R_s = {1 \over \sqrt{2 \pi}}$.
We have already rescaled the fields
$\phi_{c,s} \rightarrow \phi_{c,s} /\sqrt{2\pi}R_{c,s}$
and $\tphi_{c,s} \rightarrow \sqrt{2\pi}R_{c,s} \tphi_{c,s}$ so that
 the Hamiltonian is simply the one
for two free bosons $\phi_{c,s}$, normalized as in Eq. (\ref{efree}).
When compared with
Kane and Fisher \cite{Kane}, our
radii $R_{c,s}$ are related to their interaction strengths $g_{c,s}$
\cite{Kane} by
$\pi R_{c,s}^2=1/g_{c,s}$.

To obtain $H^P$ in $Z_P^{(a,b)}(\tq)$, we need to specify the boundary
conditions on the low energy fields.  Once again, we infer from the
finite temperature calculation of $Z_{ab}(q)$ that we impose
(anti)periodic boundary conditions on the (fermions) bosons in the
imaginary time direction.  By switching the roles of space and time,
we now compute $Z_P^{(a,b)}(\tq)$ which lead us to impose
(anti)periodic boundary conditions on $(\psi_{L,R\alpha})$
$\phi_{c,s}$ and $\tphi_{c,s}$.
By imposing antiperiodic boundary condition, $\psi_{L,R\alpha}(x) =
-\psi_{L,R\alpha}(x+L)$, we use
(\ref{ebosonizes}) to obtain the periodic boundary
conditions on $\phi_{c,s}$ and $\tphi_{c,s}$:
\begin{eqnarray}
&Q_f = \phi_f(L)-\phi_f(0)=\pi n_f R_f ~~~~
\Pi_f = \tphi_f(L)-\tphi_f(0)= {\tilde{n}_f \over 2R_f} \label{ebcbosons} \\
&{\rm where}~~f= c~{\rm or}~s
\nonumber \\
&n_c + \tilde{n}_c + n_s + \tilde{n}_s = 0 {\pmod 4}
{\rm~and~same~parity~for~all~n's}~.
\label{egluings}
\end{eqnarray}
With these boundary conditions, we can write down a mode expansion for
the bosons $\phi_{c,s}$ and $\tphi_{c,s}$ as in (\ref{emode}),
\begin{equation}
H={2\pi \over L}\sum_{f=c,s} [{1\over 4\pi}({\Ph_f}^2 + {\Qh_f}^2) +
\sum_{n=1}^\infty n({\hat m}^L_{fn} + {\hat m}^R_{fn}) - {1\over 12}]
\label{eHmodes}
\end{equation}
where the eigenvalues of $\Qh_f$ and $\Ph_f$ are $Q_f$ and
$\Pi_f$ defined in (\ref{ebcbosons})
and the $-{\pi \over 3L}$ term is the
ground state energy from (\ref{ehamilP}) for the  $c=2$ spin and
charge bosons.

The conserved Virasoro and $U_c(1) \times U_s(1)$ Kac-Moody currents
are given by
\begin{eqnarray}
T_f&=&:(\pa_+ \phi_f)^2:, ~~{\bar T_f}=:(\pa_- \phi_f)^2:
\nonumber \\
J_f&=&{1\over \pi R_f}\pa_+ \phi_f,~~{\bar J_f}=-{1\over
\pi R_f}\pa_- \phi_f
\label{ecurrents}
\end{eqnarray}
where $f=c,s$.
Expanding in modes, we obtain for $n > 0$, $f=c,s$
\begin{eqnarray}
J_{f0}&=&{1 \over 2\pi R_f}(\Ph_f + \Qh_f),~~J_{fn}=-\sqrt{n\over
\pi R_f^2}ia^L_{fn},~~J_{f-n}=J^\dagger_{fn}
\nonumber  \\
{\bar J}_{f0}&=&{1 \over 2\pi R_f}(-\Ph_f + \Qh_f),~~
{\bar J}_{fn}=\sqrt{n\over \pi
R_f^2}ia^R_{fn},~~{\bar J}_{f-n}={\bar J}^\dagger_{fn}.
\label{eJmodes}
\end{eqnarray}

We now turn to the two channel system and work with two copies of the
spin and charge bosons $\phi_f^1$ and $\phi_f^2$.  We will use the
index $f$ to denote charge $c$ and spin $s$ without further reference.
The current conservation for the two channel system at the boundary is
given by three equations,
\begin{equation}
J^1_f+J^2_f -{\bar J}^1_f-{\bar J}^2_f=0~~ {\rm and}~~
{}~~\sum_{f=c,s} T^1_f+T^2_f -{\bar T}^1_f-{\bar T}^2_f=0
\label{escconstrs}
\end{equation}
The $U(1)$ currents have to be conserved separately for charge and
spin at the boundary.  However, we only impose that the sum of spin
and
charge energy momentum be conserved.

Going to the even and odd basis,
we substitute (\ref{ecurrents}) into (\ref{escconstrs}) and define
\begin{equation}
\phi^{e,o}_f={1\over \sqrt 2}(\phi^1_f \pm \phi^2_f), \label{ebeos}
\end{equation}
giving
\begin{eqnarray}
J^e_f -{\bar J}^e_f&=&0, ~~~~ ~\sum_{f=c,s}T^e_f+T^o_f -{\bar T}^e_f
-{\bar T}^o_f=0 \nonumber \\
{\rm where}~~J^e_f=J^1_f+J^2_f &=& {\sqrt{2}\over \pi R_f}\pa_+ \phi^e_f,~~
T^e_f=:(\pa_+ \phi^e_f)^2:~~{\rm and}~~
T^o_f=:(\pa_+ \phi^o_f)^2:.
\label{econdseos}
\end{eqnarray}
$J_f^e$ are the total spin and charge currents of the two channels.
Notice that $J_f^o$ are not present in the constraints.
By combining the two current equations in (\ref{econdseos}), we deduce
that $T_c^o+T_s^o={\bar T}_c^o+{\bar T}_s^o$.  That is, in the even
channel, the boundary
preserves the
$U_c(1)\times U_s(1)$ Kac-Moody symmetries but in the odd channel, the
boundary only
preserves the smaller conformal symmetry.

To get the Ishibashi states in the even and odd basis, we need to
expand $\phi_f^{e,o}$ in modes.
We start with the mode expansions of
$\phi_f^1$ and $\phi_f^2$ as in (\ref{emode}).
By (\ref{ebeos}), we obtain mode expansion for $\phi_f^{e,o}$ as in
(\ref{emode}) with
\begin{eqnarray}
a^{e,o}_{fn} &=& {1\over \sqrt{2}}(a^1_{fn} \pm a^2_{fn})~~~~
{\rm and} \nonumber \\
\Pi^{e,o}_f&=&{1\over \sqrt{2}}(\Pi^1_f \pm \Pi^2_f)
\equiv {1\over \sqrt{2}}{\tilde{n}^{e,o}_f\over 2R_f} \nonumber \\
Q^{e,o}_f&=&{1\over \sqrt{2}}(Q^1_f \pm Q^2_f)
\equiv  {1\over \sqrt{2}}\pi R_f n^{e,o}_f
\label{etmneos}
\end{eqnarray}
where $\tn_f^{e,o}= \tn_f^1 \pm \tn_f^2$ and $n_f^{e,o}= n_f^1 \pm n_f^2$.
The gluing conditions between $\tn_f^{e,o}$ and $n_f^{e,o}$ can be derived
from (\ref{egluings}) which holds for each channel.  We obtain
\begin{eqnarray}
&\sum_{f=c,s}n^{e}_f+\tilde{n}^{e}_f=0~~({\rm mod}~4),~~~~
\sum_{i=e,o}n^{i}_s+\tilde{n}^{i}_c=0~~({\rm mod}~4) ,\\
&\sum_{i=e,o}n^{i}_s + n^{i}_c=0~~({\rm mod}~4),~~~~
\sum_{f=c,s}n^e_f+\tilde{n}^e_f+n^o_f+\tilde{n}^o_f=0 ~~
({\rm mod}~8)
\label{egluingeos}
\end{eqnarray}
and all $n's$ have the same parity.
Note that the even and odd channels are not decoupled.

The Ishibashi states for the even channel must be (\ref{eIstate})
\begin{equation}
|n^e_f=0,\tilde{n}^e_f\>>^e_I = e^{i \Pi^e_f
\phi^e_{f0}}\prod_{n=1}^\infty e^{-a^{eL \dagger}_{fn} a^{eR \dagger}_{fn}}
|0\>>
\label{eIef}
\end{equation}
because of the even $U(1)$ charge and spin symmetries at the boundary.
In the odd channel,
we have at least two choices analogous to the spinless case for each $f$:
the Ishibashi state (\ref{eIef}) with $e \rightarrow o$ which preserve
the odd $U(1)$
charge and spin current conservations or
\begin{equation}
|n^o_f,\tilde{n}^o_f=0\>>^o_I = e^{i Q^o_f
\tphi^o_{f0}}\prod_{n=1}^\infty e^{a^{oL \dagger}_{fn} a^{oR \dagger}_{fn}}
|0\>>
\label{eIof}
\end{equation}
which maximally violate the odd $U(1)$ spin and charge conservation.

With these two possibilities for the Ishibashi states in the odd channel
for each $f=c,s$, we construct four boundary states corresponding to
perfect or zero conductances for charge and spin.
Since the even Ishibashi states are the same for the four cases, we
let
\begin{equation}
|n^e_{c,s}=0,\tn^e_{c,s}\>>^e_I = |n^e_{c}=0,\tn^e_{c}\>>^e_I
\otimes |n^e_{s}=0,\tn^e_{s}\>>^e_I
\end{equation}
and the four boundary states are
\begin{eqnarray}
|1\>> &=& \sum_{\tn_{c,s}^e,\tn_{c,s}^o}{^\prime}~
|n^e_{c,s}=0,\tn^e_{c,s}\>>^e_I \otimes
|n^o_c=0,\tn^o_c\>>^o_I \otimes |n^o_s=0,\tn^o_s\>>^o_I
\sim |co,so\>>
\label{ebstatecoso}\\
|2\>> &=& \sum_{\tn_{c,s}^e,n_{c,s}^o}{^\prime}~
|n^e_{c,s}=0,\tn^e_{c,s}\>>^e_I \otimes
|n^o_c,\tn^o_c=0\>>^o_I \otimes |n^o_s,\tn^o_s=0\>>^o_I
\sim |cp,sp\>>
\label{ebstatecpsp}\\
|3\>> &=& \sum_{\tn_{c,s}^e,\tn_c^o,n_s^o}{^\prime}~
|n^e_{c,s}=0,\tn^e_{c,s}\>>^e_I \otimes
|n^o_c=0,\tn^o_c\>>^o_I \otimes |n^o_s,\tn^o_s=0\>>^o_I
\sim |co,sp\>>
\label{ebstatecosp}\\
|4\>> &=& \sum_{\tn_{c,s}^e,n_c^o,\tn_s^o}{^\prime}~
|n^e_{c,s}=0,\tn^e_{c,s}\>>^e_I \otimes
|n^o_c,\tn^o_c=0\>>^o_I \otimes |n^o_s=0,\tn^o_s\>>^o_I
\sim |cp,so\>>
\label{ebstatecpso}
\end{eqnarray}
where the primes denote summing over the quantum numbers allowed by the
gluing conditions (\ref{egluingeos}) and $|cp,so\>>$ denotes
charge-open and spin-periodic boundary state, etc..

We will give the finite size spectrums for all possible pairs of the
above boundary states.  In particular, we have worked out in the
appendix for comparison the
partition functions $Z_{(cp,sp;cp,sp)}(q)$, $Z_{(co,so;co,so)}(q)$ and
$Z_{(co,so;cp,sp)}(q)$
by directly imposing the corresponding boundary conditions.
We find that by normalizing the boundary states $|1\>>$ and $|2\>>$ to
get integer coefficients, we can recover these
partition functions.  We also normalize boundary states $|3\>>$ and
$|4\>>$ to give partition
functions from the seven other pairs of boundary states with integer
coefficients.  These are
predictions for the finite size spectrums with the corresponding pairs
of boundary conditions.

Before computing the partition functions, we need the Hamiltonian for
the $c=4$ two channel spin and charge bosons in the even and odd basis.
Following the procedure in the spinless case, we obtain
\begin{equation}
H={2\pi \over L} \sum_{i=e,o} \sum_{f=c,s}
[{1\over 4\pi}(\Ph_f^{i2} + \Qh_f^{i2}) +
\sum_{n=1}^\infty n({\hat m}_{fn}^{iL} + {\hat m}_{fn}^{iR}) - {1\over 12}]
\label{eHPeos}
\end{equation}

We find
\begin{eqnarray}
Z^{(1,1)}_P(\tq)&=&\<< 1|e^{-lH^P}|1\>> \nonumber \\
&=&
{1\over \eta (\tq)^4}\sum_{\tn_{c,s}^{e,o}}{^\prime}~
\tq^{[{1\over 64\pi}[{1\over R_c^2}(\tn_c^{e2}+\tn_c^{o2})
+ {1\over R_s^2}(\tn_s^{e2}+\tn_s^{o2})  ]}~.
\end{eqnarray}
Solving the gluing constraints (\ref{egluingeos}) by setting
$n_f^i =0$ where $i=e,o$, $f=c,s$ and letting
$\tn_f^{e,o} = \tn_f^1 \pm \tn_f^2$, we find
$\tn_c^1 = \tn_s^1 {\pmod 4}$ and $\tn_{c,s}^1$ are even and the same
conditions
for $\tn_f^2$.  It now can be written as a complete square,
\begin{eqnarray}
Z^{(1,1)}_P(\tq) &=& \{ {1\over \eta (\tq)^2}\sum_{\tn_c^1, \tn_s^1}{^\prime}~
\tq^{[ {1\over 32\pi}[{(\tn_c^1)^2 \over R_c^2}
+ {(\tn_s^1)^2\over R_s^2} ]}  \}^2 \\
&=&\{ {1 \over \eta (\tq)^2}\sum_{\tn_c+\tn_s=0 \pmod 2}
\tq^{[ {1\over 8\pi}[{\tn_c^2 \over R_c^2}
+ {\tn_s^2\over R_s^2} ]}  \}^2
\end{eqnarray}
where we have let $\tn_{c,s}^1=2\tn_{c,s}$.
Splitting the above sum into a sum of $\tn_{c,s}$ when they are both
even and odd, we obtain
\begin{eqnarray}
\{\Omega({1\over {\pi R_c^2}},0,0;\tq) \Omega({1\over {\pi
R_s^2}},0,0;\tq) +
\{\Omega({1\over {\pi R_c^2}},0,\pi;\tq) \Omega({1\over {\pi
R_s^2}},0,\pi;\tq) \} ^2 \\
= \pi^2 R_c^2 R_s^2\{\Omega({\pi R_c^2},0,0;q)
\Omega({\pi R_s^2},0,0;q) +
\{\Omega({\pi R_c^2},\pi,0;q) \Omega({\pi
R_s^2},\pi,0;q) \} ^2
\end{eqnarray}
where we have used (\ref{eOmodtmn}).  It can then be rewritten as
\begin{eqnarray}
(2\pi R_c R_s)^2 \{ {1\over \eta (q)^2}\sum_{n_c+n_s=0 \pmod 2}
q^{{1 \over 2} \pi R_c^2 n_c^2 + {1 \over 2}\pi R_s^2 n_s^2} \}^2
\equiv {\cal Z}_{(1,1)} (q)
\label{epartcoso}
\end{eqnarray}
Going through the same steps as before to solve the gluing constraints
then modular transforming, we find
\begin{equation}
Z_P^{(2,2)} (\tq) =  {1\over \eta (q)^4}\sum{^\prime}~
q^{ [{1 \over 4} \pi R_c^2 n_c^2 + {\tn_c^2 \over {16\pi R_c^2}} +
{1 \over 4}\pi R_s^2 n_s^2 + {\tn_s^2 \over {16\pi R_s^2}} ] }
\equiv {\cal Z}_{(2,2)} (q)
\label{epartcpsp}
\end{equation}
where the gluing conditions in this sum  are $n_c +n_s +\tn_c +
\tn_s=0 {\pmod 4}$
and all the $n's$ have the same parity.
Just like the spinless case, this partition function is modular
invariant because the boundary conditions on the bosons are periodic
(fermions $\psi_{L,R\alpha}$ are antiperiodic) in both the time and
space directions.
We also find from the boundary states that
\begin{equation}
Z_P^{(1,2)} (\tq) =2\pi R_c R_s  { 1\over \eta (q)^2 }\sum_{n_c+n_s=0 \pmod 2}
q^{{1 \over 4} \pi R_c^2 n_c^2 + {1 \over 4}\pi R_s^2 n_s^2}
W_+ (q)^2 \equiv {\cal Z}_{(1,2)}(q)
\label{ezcosocpsp}
\end{equation}
For these to be consistent partition functions, we need to normalize away
the noninteger coefficients.  Normalizing the boundary states by
\begin{equation}
|co,so\>> = {1\over 2\pi R_c R_s}|1\>> ~~{\rm and}~~
|cp,sp\>> = |2\>> ,
\label{enorm}
\end{equation}
we reproduce the same spectrum as the ones from the appendix.

We predict the following partition functions:
\begin{eqnarray}
Z_P^{(3,3)} (\tq) &=& \pi R_c^2 {1\over \eta (q)^4}
\sum_{m=n=k+l \pmod 2}q^{ [{m^2 \over {16\pi R_s^2}} +
{1 \over 4} \pi R_s^2 n^2 + {1 \over 2}\pi R_c^2 (k^2+l^2)]}
\label{epartcosp} \\
Z_P^{(1,3)} (\tq) &=& 2\pi R_c R_s\sqrt{\pi R_c^2}{1\over \eta (q)^3}
\sum_{k+m+n=0 \pmod 2}q^{ [{1 \over 4} \pi R_s^2 k^2 +
{1 \over 2}\pi R_c^2 (m^2+n^2)]}W_+ (q) \\
Z_P^{(2,3)} (\tq) &=& \sqrt{\pi R_c^2} {1\over \eta (q)^3}
\sum_{m=n=k \pmod 2}q^{ [{n^2 \over {16\pi R_s^2}} +
{1 \over 4} \pi R_s^2 m^2 + {1 \over 4}\pi R_c^2 k^2]}W_+ (q)
\end{eqnarray}
By exchanging charge and spin, we obtain the other three spectra
$Z_P^{(4,4)}$, $Z_P^{(1,4)}$ and $Z_P^{(2,4)}$.
Finally, we showed
\begin{equation}
Z_P^{(3,4)} = Z_P^{(1,2)} \label{ezmixed}
\end{equation}
We find that if we normalized the boundary states with (\ref{enorm}) and
\begin{eqnarray}
|co,sp\>> = {1\over \sqrt{\pi R_c^2}}|3\>> ~~{\rm and}~~
|cp,so\>> = {1\over \sqrt{\pi R_s^2}}|4\>>
\label{enormop}
\end{eqnarray}
then all the partition functions will have integer coefficients.
With the normalizations (\ref{enorm}) and (\ref{enormop}), we see from
(\ref{ezmixed}) that the ground state is two-fold degenerate with
charge-open spin-periodic boundary condition at one end and charge
periodic spin open boundary condition at the other.

The spin and charge conductances are defined as in (\ref{econduct}) where the
currents are the spin or charge currents in (\ref{eJmodes}).
We find that the spin and charge conductances are zero or
$G_f = {e^2 / 2\pi R^2_f h}$ depending on the boundary state
(\ref{ebstatecoso})-(\ref{ebstatecpso}) we use.
That is, for the open and periodic boundaries, we get zero and perfect
conductances respectively.

\subsection{operator content and ground state degeneracies}
Once again, we can determine the operator content from the finite size
spectrum with the
same boundary conditions at both ends.
For these operators to enter the interaction Hamiltonian, they must
have all the symmetry properties  of the system.  In particular, the
operator must be real, $U(1)$ spin and charge invariant, and other
additional symmetries that we wish to impose like parity, etc..

Let us first discuss the $U_c(1)\times U_s(1)$ symmetries.
The appropriate transformations are
\begin{equation}
\psi_{\uparrow,\downarrow}
\stackrel{U_c(1)}{\rightarrow} e^{-i\alpha_c}\psi_{\uparrow,\downarrow}
{}~~{\rm and}~~
\psi_{\uparrow,\downarrow}
\stackrel{U_s(1)}{\rightarrow} e^{\mp i\alpha_s}\psi_{\uparrow,\downarrow}
\end{equation}
The $U_c(1)$ transformation is spin blind and the generator of
$U_s(1)$ rotation about the $z$-axis is $e^{(-i\alpha_s \sigma_z)}$ acting
on the two component spinor $\psi_{\uparrow,\downarrow}$.
{}From (\ref{ebosonizes}), we see that these transformations can be achieved by
\begin{eqnarray}
\phi_c &\rightarrow &\phi_c  ~~~~~~~~~~~~~\phi_s \rightarrow
\phi_s  \nonumber \\
\tphi_c &\rightarrow &\tphi_c + {\alpha_c \over \pi R_c}~~~~
\tphi_s \rightarrow \tphi_s + {\alpha_s \over \pi R_s}
\label{ecsbtmn}
\end{eqnarray}
Using $\phi_{c,s}=\phi_{c,s}^L + \phi_{c,s}^R$ and $\tphi_{c,s}=\phi_{c,s}^L -
\phi_{c,s}^R$, we find
\begin{equation}
\phi_f^L \rightarrow \phi_f^L + {\alpha_f \over 2\pi R_f}
{}~~{\rm and}~~
\phi_f^R \rightarrow \phi_f^R - {\alpha_f \over 2\pi R_f}.
\label{ecsbLtmn}
\end{equation}

{}From the charge and spin-open partition function (\ref{epartcoso}),
we see that the lowest dimensional operators have dimensions
$0,~{\pi R_c^2 \over 2}+ {\pi R_s^2 \over 2},~ 2\pi R_c^2,~2\pi R_s^2$,
etc..   In the two channels pure left moving interpretation, they
correspond to the operators the identity,
$e^{ \pm 2\pi iR_c\phi^L_c \pm 2\pi iR_s\phi^L_s}$,
$e^{ \pm 4\pi iR_c\phi^L_c}$, $e^{ \pm 4\pi iR_s\phi^L_s}$,etc..
The two decoupled channels can interact with each other via these
boundary operators.  Preserving the $U_{c,s}(1)$ symmetries, the
lowest dimension candidates are
$e^{ \pm 2\pi iR_c(\phi^{L1}_c-\phi^{L2}_c)
\pm 2\pi iR_s(\phi^{L1}_s - \phi^{L2}_s) }$,
$e^{ \pm 4\pi iR_c(\phi^{L1}_c-\phi^{L2}_c)}$ and
$e^{ \pm 4\pi iR_s(\phi^{L1}_s-\phi^{L2}_s)}$.
For simplicity, we will proceed with the symmetry $S_i^z \rightarrow
-S_i^z$.
In the fermion language, they correspond to
hopping of one fermion between the two channels
\begin{equation}
t_e~(\psi^\dagger_{L\uparrow 1} \psi_{L\uparrow 2}+
\psi^\dagger_{L\downarrow 1} \psi_{L\downarrow 2})~+~{\rm h.c.}
\sim cos2\pi R_s(\phi_{s1}^L-\phi_{s2}^L)
[t_e~e^{i2\pi R_c(\phi_{c1}^L-\phi_{c2}^L}) +~{\rm h.c.}], \label{ete}
\end{equation}
hopping of a charge two spin singlet,
\begin{equation}
t_c~(\psi^\dagger_{L\uparrow} \psi^\dagger_{L\downarrow})_1 ~
(\psi_{L\uparrow} \psi_{L\downarrow})_2 ~+~{\rm h.c.}
\sim t_c~e^{i4\pi R_c(\phi_{c1}^L-\phi_{c2}^L)} +~{\rm h.c.}\label{etc}
\end{equation}
and
hopping of a neutral spin one object
\begin{equation}
t_s~(\psi^\dagger_{L\uparrow}\psi_{L\downarrow})_1 ~
(\psi^\dagger_{L\downarrow} \psi_{L\uparrow})_2 ~+~{h.c}
\sim t_s~e^{i4\pi R_s(\phi_{s1}^L-\phi_{s2}^L)}+~{\rm h.c.}. \label{ets}
\end{equation}
The stability of the open fixed point is governed by the relevance of
these operators.  For it to be stable, all operators must have
dimensions greater than one.  That is
$\pi R_c^2+\pi R_s^2 > 1$ and $4\pi R_{c,s}>1$.

{}From the charge and spin-periodic partition function
(\ref{epartcpsp}), the lowest
$U_{c,s}(1)$ invariant operators are the identity,
$e^{\pm i\phi_c/R_c \pm i \phi_s/R_s}$,
$e^{\pm 2i\phi_c/R_c}$ and $e^{\pm 2i\phi_s/R_s}$ with dimensions
$0$, ${1/4\pi R_c^2 + 1/4\pi R_s^2}$, ${1/\pi R_c^2}$ and ${1/\pi
R_s^2}$.
In the fermion language, they correspond to the backscattering of a
fermion
\begin{equation}
v_e~(\psi^\dagger_{L\uparrow} \psi_{R\uparrow}+
\psi^\dagger_{L\downarrow} \psi_{R\downarrow}) ~+~{\rm h.c.}
\sim [v_e~e^{i\phi_c/R_c}+ {\rm h.c.}]cos(\phi_s/R_s),
\label{eve}
\end{equation}
backscattering of a charge two spin singlet object
\begin{equation}
v_c~\psi^\dagger_{L\uparrow} \psi^\dagger_{L\downarrow}~
\psi_{R\uparrow} \psi_{R\downarrow}~+~{\rm h.c.}
\sim v_c~e^{i2\phi_c/R_c}+ {\rm h.c.} \label{evc}
\end{equation}
and backscattering of a spin
one charge neutral object
\begin{equation}
v_s~\psi^\dagger_{L\uparrow}\psi_{L\downarrow}~
\psi^\dagger_{R\downarrow} \psi_{R\uparrow} ~+~{\rm h.c.}
\sim v_s~e^{i2\phi_s/R_s} ~+~{\rm h.c.}\label{evs}
\end{equation}
respectively.
For the charge and spin-periodic fixed point to be stable,
the above operators must be irrelevant with dimensions greater than one.

About the charge-open and spin-periodic fixed point, we see from
(\ref{epartcosp}) that the lowest dimensions of the operators are $0$,
$1/4\pi R_s^2$, $1/\pi R_s^2$ and $\pi R_c^2$.  They correspond to the identity
operator, $e^{\pm i\phi_s/R_s}$, $e^{\pm 2i\phi_s/R_s}$ and $e^{\pm
2\pi i R_c(\phi_{c1}^L - \phi_{c2}^L)}$ since we expect that the spin
field $\phi_s$ to be periodic
across the boundary but not the charge $\phi_c$.
In the fermion language, they correspond to backscattering of a
fermion $v_e$  (\ref{eve}) which in
this case reduces to backscattering of spin,
backscattering of a spin one charge neutral object $v_s$ (\ref{evs})
and the hopping
of a fermion across the impurity
$t_e$ (\ref{ete}) which
reduces to hopping of charge.  Essentially the operators $e^{i\phi_c(0)/R_c}$
and $e^{i2\pi R_s[\phi_{s1}^L(0)-\phi_{s2}^L(0)]}$ develop non-zero expectation
values in this phase because we expect $|v_c|$ and $|t_s|$ to be
infinite at this fixed point.  This reduces the dimension of the
$t_e$ and $v_e$
operators. The stability of this fixed point is governed by the irrelevance of
the lowest dimensional operators $t_e$ and $v_e$.  We note that the $t_e$
operator was overlooked in Ref. \cite{Kane}.  We further note, following Kane
and Fisher, that, if we impose parity, all these $t$ and $v$ parameters become
real.  Choosing a particular sign for $v_c$ then leads to
$<e^{i\phi_c(0)/R_c}>=0$ so that the $v_e$ term vanishes.\footnote{Choosing
a sign for $v_c$ can be seen as choosing a microscopic model.}
Similarly, with
parity, an appropriate choice of the sign of $t_s$ causes the $t_e$ term to
vanish.

The results of a similar analysis of the charge-periodic spin-open
fixed point can be obtained by
exchanging charge and spin in the above case.
The stability of the four fixed points are shown in figure 2, in the generic
case where the $v_e$ and $t_e$ operators are non-vanishing in the mixed phase.

The ground state degeneracies also give {\it relative} stability for these
four fixed points.  The one fixed point that is stable has its ground
state degeneracy smaller than the rest.
The ground state degeneracies for the four boundary states are
precisely the respective normalization coefficients in (\ref{enorm})
and (\ref{enormop}).
We obtain a phase diagram
very similar to the results of the operator content analysis as shown
in figure 3.  If
there were no nontrivial unstable fixed points with higher ground
state degeneracies, then this would be the correct
phase diagram.

\subsection{Resonant tunneling}
By using the operator contents of the various boundary fixed points,
we can find out the stability of the fixed points as a function of
$R_c$ and $R_s$.

About the charge and spin open boundary, the lowest dimensional
operators that can couple
to the impurity are
$e^{ \pm 2\pi iR_c\phi^L_c \pm 2\pi iR_s\phi^L_s}$,
$e^{ \pm 4\pi iR_c\phi^L_c}$ and $e^{ \pm 4\pi iR_s\phi^L_s}$
with dimensions
${\pi R_c^2 \over 2}+ {\pi R_s^2 \over 2},~ 2\pi R_c^2$ and $2\pi
R_s^2$.
For instance, the first operator arises from
$[\psi_{L\uparrow 1}^\dagger \eta_{I\uparrow} + \hbox{h.c.}]$ and the
second and third arise from
$(\psi_{L\uparrow}^\dagger \psi_{L\downarrow}^\dagger)
\eta_{I\uparrow}\eta_{I\downarrow}$ and
$(\psi_{L\uparrow}^\dagger \psi_{L\downarrow})
\eta_{I\uparrow}^\dagger\eta_{I\downarrow}$.
The stability of this fixed point is governed by the condition that all
these operators have dimensions greater than unity.

About the periodic case, we fine tune away the backscattering of a
single electron corresponding to the operator $v_e$.
For a symmetric potential, we only need to fine tune
one parameter, for instance the chemical potential at the impurity
site.  For an asymmetric potential, we need to fine tune two
parameters to eliminate this backscattering term to achieve resonance.
After fine tuning away this backscattering term,
The next two lowest dimensional operators are $v_c$ and $v_s$ with
dimensions ${1/\pi R_c^2}$ and ${1/\pi R_s^2}$.

About the charge-open spin-periodic fixed point, we continue to fine
tune $v_e$ to zero.
{}From the partition function (\ref{epartcosp}) we find those
operators, either by themselves or when coupled to the impurity
$\eta_{I\alpha}$, that
are not eliminated by the various symmetries.
The lowest dimensional operators allowed by the symmetries are $v_s$
(\ref{evs}) and the hopping operator
that connects the chain to the resonant site
$[\psi^{\alpha \dagger}_1\eta_{I\alpha} +
\hbox{h.c.}]$.
We conclude that the stability of this fixed point is
now determined by the irrelevance of
these two operators, that is,  $1/\pi R_s^2>1$ and
$\pi R_c^2 /2 + 1/16\pi R_s^2 + \pi R_s^2/4 >1$.
By exchanging charge and spin, we obtain similar
conclusions for the spin-open charge-periodic fixed point.

The relative stability of these resonant fixed points determined from
their ground state degeneracies give
a phase diagram same as the one about the charge and spin periodic
fixed point.
\section{Conclusions}
We have shown that at low energies,
interacting fermions coupled to a local potential or an impurity can
be turned into a \bcp\ problem.
In the spinful case we are able to analyze  the charge open and spin
periodic boundary fixed point in a somewhat more systematic way than in
 \cite{Furusaki},
\cite{Kane}.  We agree completely with these papers after correcting
a minor error concerning the stability of this charge open and spin
periodic fixed point.

A possible way of finding the nontrivial fixed points of \cite{Kane}
is to guess a
boundary state that by construction partially conducts and compute
the partition functions by taking all matrix elements with the known
boundary states.  When modular transformed, we require that these
partition functions must have integer coefficients. So far, we have failed
to guess such a state. Another way to proceed  may be to fix the
radius of interaction to a rational value in the region where we
expect nontrivial fixed points and then use fusion
to go from a trivial fixed point to  a nontrivial one.

\acknowledgments
We would like to thank J. Gan,  M.P.A. Fisher, C. Kane,  A.W.W. Ludwig and
J. Sagi   for useful discussions.
This research was supported in part
by NSERC of Canada.

\appendix
\section{}
We wish to derive (\ref{emodtmn}) in this appendix.
Substituting the Poisson sum formula
\begin{equation}
\sum_{Q=-\infty}^\infty f(Q) = \int dx \sum_{P=-\infty}^\infty
e^{2\pi i P x} f(x)
\end{equation}
and (\ref{echaracter}) into the left hand side of (\ref{emodtmn}), we
obtain
\begin{eqnarray}
\sum_{Q=-\infty}^\infty e^{i\delta_1 Q} \chi_Q(a,\delta_2;q) =
{1\over \eta (q)} \int dx \sum_{P=-\infty}^\infty e^{2\pi i P x}
e^{i \delta_1 x} q^{{a \over 2}(x - {\delta_2 \over {2\pi }})^2}
\end{eqnarray}
Shifting $x$ to
$x+{\delta_2 \over 2\pi }$ and using
$q = e^{-\pi {\beta \over l}} $,
the above becomes
\begin{eqnarray}
& {1\over \eta(q)} e^{{i \delta_1 \delta_2} \over {2\pi}}
\sum_{P=-\infty}^\infty  e^{i P \delta_2} \int dx~
e^{- \pi a {\beta \over {2l}} [x^2 - {{i4l} \over {\beta a}}
(P + {\delta_1 \over {2\pi}}) x ]}
\\
=& {1\over \eta(q)} e^{{i \delta_1 \delta_2} \over {2\pi}}
\sum_{P=-\infty}^\infty  e^{i P \delta_2} \sqrt{{2l}\over{a \beta}}
e^{-{{2l \pi}\over {\beta a}} (P + {{\delta_1} \over {2\pi}})^2}.
\end{eqnarray}
Using the fact that under a modular transform, the Dedekind function
transform as
$\eta (q) = \sqrt{2 l \over \beta} \eta (\tq)$ \cite{Ginsparg}, we get
\begin{eqnarray}
&e^{{i \delta_1 \delta_2} \over {2\pi}} {1 \over \sqrt{a}}
{1 \over {\eta (\tq)}} \sum_{P=-\infty}^\infty  e^{i P \delta_2}
\tq^{{1 \over 2a}(P+{\delta_1 \over {2\pi}})^2}
\\
=&e^{{i \delta_1 \delta_2} \over {2\pi}} {1 \over \sqrt{a}}
\sum_{P=-\infty}^\infty  e^{i P \delta_2}
\chi_P ({1 \over a}, -\delta_1; \tq)
\end{eqnarray}
which is the right hand side of (\ref{emodtmn}).

\section{}
In this appendix, we derive the finite size spectrums for the spinless
fermions by directly imposing the periodic and open boundary
conditions.
We will choose appropriate boundary conditions at $x=0, \pm 1$ for the
one channel system of length $2l$ to give the finite size spectra of
the two channel folded system of length $l$ with periodic or open
boundary conditions placed at $x=0,l$.  We choose, $2 k_F l = 2 \pi (N + {1
\over 2})$.

We now derive the finite size spectrum for the two channel
system of length $l$ with
periodic-periodic boundary conditions at the two ends.
Unfolding this  into a one channel system, we have
a periodic system of length $2l$.
For the one channel periodic system on $2l$, we simply obtain its
spectrum from (\ref{eHmode}) with the periodic quantization conditions
(\ref{ebcboson}).
Substituting $L=2l$ and (\ref{ebcboson}) into (\ref{eHmode}), we obtain
\begin{equation}
E_{P}={\pi \over l} [{1 \over 4\pi} ({m^2 \over {4 R^2}} +
4 \pi^2 n^2 R^2) + \sum_{n=1} n (m_n^L + m_n^R) - {1 \over 12}]~.
\end{equation}
By (\ref{epartab}), we have
\begin{equation}
Z_{pp}(q)=\sum_{m,n}{^\prime}~ \sum_{m_1^{L,R},m_2^{L,R},...}
e^{-\beta E_{P}}
\end{equation}
where the gluing constraint is (\ref{egluing}).  Splitting the sum
into when both $m=2k$ and $n=2p$ are even and when both $m=2k+1$ and
$n=2p+1$ are odd,
we obtain for $k,p \in {\bf I}$
\begin{eqnarray}
Z_{pp}(q)&=&{1\over \eta (q)^2} \sum_{m,n}{^\prime}~
q^{{m^2\over 16\pi R^2}+n^2\pi R^2}  \\
&=& {1 \over \eta ^2} \sum_{k,p} q^{{k^2 \over {4 \pi R^2}} +
{4 \pi p^2 R^2}} + {1 \over \eta^2} \sum_{k,p} q^{{{(k +{ 1 \over 2})}^2
\over {4 \pi R^2}} +
{4 \pi {(p + {1 \over 2})}^2 R^2}} \\
&=& \Omega ({1\over 2\pi R^2},0,0;q)  \Omega (8\pi
R^2,0,0;q)+ \Omega ({1\over 2\pi R^2},0,\pi;q) \Omega (8\pi R^2,0,\pi;q),
\end{eqnarray}
in agreement with (\ref{ezpp}).

We now consider the finite size spectrum for the two channel open
boundaries system.  This system is equivalent to two independent copies
of the one channel open boundary system.  We will first work out the finite
size spectrum for the one channel case.
For open boundary conditions, we impose at the boundaries
$\pa_x \psi (0) = \pa_x \psi (l) = 0$
for the one channel case.
Using (\ref{ekF})
and the fact that the low energy modes have $k<<k_F$, we obtain
$\psi_L (0) = \psi_R (0)$ and $ \psi_L(l) = e^{2ik_Fl} \psi_R(l)$.
Bosonizing and setting $2 k_F l = 2 \pi (N + {1 \over 2})$
we obtain
$$\phi (0) = 0~~{\rm and}~~ \phi (l) = 2 \pi n R$$
where we took
\begin{equation}
[ \phi_L (x), \phi_R (y)] =\left\{
\begin{array}{ll}
0          & x,y=0   \\
i\over 4   & 0<x,y<l \\
i\over 2   & x,y=l
\end{array}
\right .
\label{ecommu}
\end{equation}
The boundary conditions in $\phi$ and the commutation relation are
compatible with the mode expansion
\begin{equation}
\phi (x,t) = \hat{Q} {x \over l} + \sum_{n=1}^\infty
{ 1 \over {\sqrt{\pi n}}} \sin {{\pi n x} \over l}
[e^{-{{i \pi n t} \over l}} a_n + h.c.]
\label{ephiomode}
\end{equation}
where eigenvalues of $ \hat{Q}$ is $ Q = 2 \pi n R $.
The boundary condition restricts the zero mode
$\phi_0=\phi^L_0+\phi^R_0=0$ but
$\tphi_0=\phi^L_0-\phi^R_0$ is nonzero and is conjugate to $\hat{Q}$.
Using (\ref{efree}),
we obtain
\begin{equation}
E = {\pi \over l} [ {1 \over {2\pi}}Q^2 +  \sum_{n=1}^\infty n m_n
-{1\over 24}]
\label{espectoo}
\end{equation}
where ${m_n}$ is the eigenvalue of
$a_n^\dagger a_n$.
For one channel, the partition function is
\begin{eqnarray}
Z_{oo}^1(q) &=& \sum_Q \sum_{m_1,m_2 \ldots}
e^{-{{\beta \pi} \over l} [{1 \over {2\pi}} Q^2 +
\sum_{p=1}^\infty p m_p -{1\over 24}]} \\
&=& {1 \over {\eta (q)}} \sum_n q^{2 \pi n^2 R^2} \\
&=& \Omega ( 4 \pi R^2, 0, 0; q).
\end{eqnarray}
For two channels with open boundaries, we simply have two uncoupled
copies of the one
channel problem. Therefore the partition function for the two channel
open boundaries system is
$$Z_{oo}(q) = \Omega {(4 \pi R^2, 0, 0; q)}^2~,$$
in agreement with (\ref{ezoo}) after normalization (\ref{enbstates}).

For the two channel periodic-open case, we equivalently
have periodic
boundary condition at $x=0$ and open boundary conditions
at $x= \pm l$ in the unfolded one channel system.
In other words, we have open
boundary
conditions at $x= \pm l$ for the one channel system of length $2l$.
Therefore, the spectrum is
(\ref{espectoo}) with $l \rightarrow 2l$
and the partition function is
\begin{eqnarray}
Z_{op}(q) = q^{-{1\over 48}}\sum_n \sum_{m_1,m_2 \ldots}
q^{\pi R^2 n^2 + {1 \over 2} \sum_{p=1}^\infty p m_p}
\label{epartmix}
\end{eqnarray}
We split the descendents ${1 \over 2} \sum_{p=1}^\infty p m_p=
\sum_{p=1}^\infty p m_{2p}+\sum_{p=1}^\infty (p-{1\over 2}) m_{2p-1}$
and then sum over the $m_{2p}$ to get the Dedekind function.
Then,
\begin{eqnarray}
Z_{op}(q) &=& \Omega ( 2\pi R^2, 0, 0; q)~  q^{1\over 48}
\sum_{m_1,m_3 \ldots} \prod_{p=1}^\infty q^{ (p-{1\over 2}) m_{2p-1}}
\nonumber \\
&=&\Omega ( 2\pi R^2, 0, 0; q)~ q^{1\over 48}
\prod_{p=1}^\infty {1\over 1-q^{p-{1\over 2}}}~. \nonumber
\end{eqnarray}
Using Euler's identity $\prod_{n=1}^\infty (1-x^{2n-1})(1+x^n)=1$
\cite{Ginsparg} with $x=\sqrt{q}$ and
extracting out a Dedekind function, we
rewrite
\begin{eqnarray}
 q^{1\over 48} \prod_{p=1}^\infty {1\over 1-q^{p-{1\over 2}}}
&=& q^{1\over 16} {1\over \eta (q)} \prod_{p=1}^\infty (1-q^p)(1+q^{p
\over 2}) \nonumber \\
&=& q^{1\over 16} {1\over 2\eta (q)} \prod_{p=1}^\infty
(1-q^{p\over 2})(1+q^{p\over 2}) (1+q^{{p\over 2}-{1\over 2}}).
\end{eqnarray}
We now use the Jacobi triple product identity \cite{Ginsparg}
$$ \prod_{n=1}^\infty  (1-x^{2n})(1+yx^{2n-1}) (1+y^{-1}x^{2n-1})
=\sum_{k=-\infty}^{\infty}y^k x^{k^2}$$
with $x=y=q^{1/4}$
and turn the above product into a sum.
We finally arrive at
\begin{equation}
Z_{op}(q) =
\Omega ( 2\pi R^2, 0, 0; q) W_+(q)
\label{epartsop}
\end{equation}
where $W_+(q)$ is given by (\ref{ewp}).

\section{}
In this appendix, we derive the finite size spectrums for the spin
${1\over 2}$ fermions by imposing periodic and open boundary conditions.
For a two channel periodic-periodic system of length $l$, it is
equivalent to a one channel periodic system of length $2l$.  We obtain
the finite size spectrum by substituting (\ref{ebcbosons}) and $L=2l$
into (\ref{eHmodes}), giving
$$
E_p={2\pi \over 2l}\sum_{f=c,s} [{1\over 4\pi}
({\tn_f^2\over 4R_f^2} + \pi^2 n_f^2 R_f^2) +
\sum_{n=1}^\infty n(m^L_{fn} + m^R_{fn}) - {1\over 12}]
$$
Therefore, the partition function for this periodic-periodic system is
\begin{eqnarray}
Z_{pp}(q)&=&\sum_{\tn_{f},n_{f}}{^\prime}~\sum_{m^{L,R}_{fn}}
e^{-\beta E_p} \nonumber \\
&=& {1\over \eta (q)^4} \sum_{\tn_{f},n_{f}}{^\prime}~
q^{[{1\over 4}\pi n_f^2 R_f^2 + {\tn_f^2 \over 16 \pi R_f^2}]}
\end{eqnarray}
where $n_{c,s}$ and $\tn_{c,s}$ obey the gluing conditions
(\ref{egluings}).  This agrees with (\ref{epartcpsp}).

Consider now the two channel open-open case.  This decouples into two
one channel system each of length $l$.
For each of the  one channel system, we again impose boundary
conditions
$\pa_x\psi_\alpha (0)= \pa_x \psi_\alpha (l) =0$ leading to
$\psi_{L \alpha}(0)=\psi_{R \alpha}(0)$ and
$\psi_{L \alpha}(l)=e^{2ik_Fl} \psi_{R \alpha}(l)$ when we use
(\ref{ekFs}).  We then bosonize these boundary  conditions with
(\ref{ebosonizes}) to obtain boundary conditions on the bosons and
just like the spinless case, we use the commutation relations
(\ref{ecommu}) for the charge
and spin bosons.  We get
$$
( {\phi_c \over R_c} \pm {\phi_s \over R_s}) (0)=0 ~~{\rm and}~~
( {\phi_c \over R_c} \pm {\phi_s \over R_s}) (l)=2\pi n_{\pm}
$$
where $n_{\pm} \in {\bf I}$ and we have chosen $2k_Fl= 2\pi (N+{1\over
2})$. It follows that
$$
\phi_{c,s}(0)=0 ~~{\rm and}~~ \phi_{c,s}(l)=\pi R_{c,s}n_{c,s}
$$
where $n_{c,s}=n_+ \pm n_-$ or equivalently we simply impose
gluing conditions
$n_c=n_s {\pmod 2}$.
Expanding $\phi_{c,s}$ in modes compatible with the above boundary
conditions as in (\ref{ephiomode}), we obtain for
$Q_{c,s}= \phi_{c,s}(l) -\phi_{c,s}(0)=\pi R_{c,s} n_{c,s}$,
\begin{equation}
E_o = {\pi \over l}[{1\over 2\pi}(Q_c^2 + Q_s^2)
+\sum_{p=1}^\infty p(m_{cp}+m_{sp}) -{1\over 12}].
\label{espectsoo}
\end{equation}
Hence for a one channel system with open boundaries, the partition
function is
\begin{eqnarray}
Z_{oo}^1 (q)&=& \sum_{n_c=n_s \pmod 2} \sum_{\{m_{cp},m_{sp}\}} e^{-\beta
E_o} \nonumber \\
&=& {1\over \eta(q)^2} \sum_{n_c=n_s \pmod 2}
q^{[{\pi \over 2}n_c^2 R_c^2 +{\pi \over 2}n_s^2 R_s^2]}
\end{eqnarray}
Hence the partition function for the two channel open boundaries
system is
$$Z_{oo}(q)= (Z_{oo}^1)^2~,$$
which agrees with (\ref{epartcoso}) after normalization (\ref{enorm}).

To obtain the finite size spectrum of the two channel periodic-open
system,
we again use the fact that it is
equivalent to a one channel system of length $2l$ with open boundaries.
The spectrum is therefore
(\ref{espectsoo}) with $l \rightarrow 2l$ and the partition function
becomes
\begin{equation}
Z_{op}(q) = q^{-{1\over 24}} \sum_{n_c=n_s \pmod 2}
\sum_{\{ m_{cp},m_{sp} \}}
q^{[{\pi \over 4}n_c^2 R_c^2 +{\pi \over 4}n_s^2 R_s^2 +{1\over 2}
\sum_{p=1}^\infty p(m_{cp}+m_{sp}) ]}
\end{equation}
Following the steps in the spinless case from (\ref{epartmix}) to
(\ref{epartsop}) for both the charge and spin descendents, we obtain
\begin{equation}
Z_{op}(q)={1\over \eta(q)^2} \sum_{n_c=n_s \pmod 2}
q^{[{\pi \over 4}n_c^2 R_c^2 +{\pi \over 4}n_s^2 R_s^2]}~ W_+(q)^2~,
\end{equation}
in agreement with (\ref{ezcosocpsp}).

\figure{Figure 1ab: We fold the system of length $2L$ into a two-channel
system of
length $L$ after we identify the boundary conditions at $-L$ and $L$. }

\figure{Figure 2: The phase diagram in the space of the charge and spin
interaction strength $R_c$ and $R_s$ without imposing parity
invariance. (cp,sp), (co,so), (co,sp)
and (cp,so) denote the four stable boundary fixed points: charge (c)
and spin (s)
periodic (p) or open (o).
The unshaded region is where both  (cp,sp) and (co,so) fixed
points are stable.  An unstable fixed point should
separate these stable phases.}

\figure{Figure 3: The phase diagram according to the ground state
degeneracies.
Since the unstable fixed point is expected to have a higher ground
state degeneracy than the stable phases, it does not show up when we
only compare the relative stability of the four phases.}

\end{document}